# Triplet-Polaron Interaction Induced Upconversion from Triplet to Singlet: a New Way to Obtain Highly Efficient OLEDs

*Ablikim Obolda,[†] Qiming Peng,[†] Chuanyou He, [†] Tian Zhang, Jiajun Ren, Hongwei Ma, Zhigang Shuai\* and Feng Li\**

Prof. Dr. Feng Li, Dr. Ablikim Obolda, Dr. Qiming Peng, MS. Chuanyou He, Dr. Hongwei Ma
State Key Laboratory of Supramolecular Structure and Materials, Institute of Theoretical Chemistry, Jilin University, Qianjin Avenue, Changchun, 130012, P. R. China
E-mail: *lifeng01@jlu.edu.cn*
Prof. Dr. Zhigang Shuai, Dr. Tian Zhang, Dr. Jiajun Ren
Key Laboratory of Organic OptoElectronics and Molecular Engineering, Department of Chemistry, Tsinghua University, Beijing 100084, P. R. China
E-mail: *zgshuai@tsinghua.edu.cn*
[†]     these authors contributed equally to this work.

**Abstract**：The triplet harvesting is a main challenge in organic light-emitting devices (OLEDs), due to the radiative decay of triplet is spin-forbidden. Here, we designed and synthesized two D-A type molecules, TPA-TAZ and TCP. The OLEDs based on them exhibit deep-blue emission and the singlet formation ratios are higher than the simple spin-statistics of 25 %. Specially, a TPA-TAZ-based OLED achieves a maximum EQE of 6.8 %, which is the largest value of the undoped OLEDs with CIE(y)< 0.06 (the EBU blue standard) up to date. Comprehensive experiments eliminate the triplet-harvesting processes of thermally activated delayed fluorescence and triplet-triplet annihilation. Instead, the triplet-polaron interaction induced upconversion from triplet to singlet through one-electron transfer mechanism is proposed, and proven by the magneto-current measurement and quantum chemistry computation. Our results may offer a new route to break through the 25 % upper limit of IQE of fluorescent OLEDs, especially, the deep-blue fluorescent OLEDs.

Since the milestone work of Tang *et al.* in 1987, organic light-emitting diodes (**OLEDs**) have attracted extensive attention due to their wide applications in the full color flat-panel displays and solid-state lighting sources.[1] In an OLED, the singlet to triplet exciton formation ratio is expected to be 1:3, according to the simple spin statistics.[2] Generally, only singlet can radiatively decay, and the triplets which account for 75 % of the total excitons are wasted. Hence, plenty of studies have been focused on harvesting the triplet excitons. To realize this goal, phosphorescent materials were developed and have achieved great success, as they can approach 100 % internal quantum efficiency (**IQE**) by harvesting both singlet and triplet excitons.[1c-1d,3] However, practically useful phosphorescent materials are concentrated to the expensive Ir and Pt complexes, and satisfactory deep-blue phosphorescent OLEDs are hardly obtained.[4] Thus, fluorescent emitters are commonly used for deep-blue OLEDs.

For harvesting triplet excitons in a fluorescent OLED, triplet-triplet annihilation (**TTA**) is a feasible method.[5] In this process, one singlet can be generated by consuming two triplets, leading to an up-limit IQE of 62.5 %.[5c-5d] Another strategy to make use of the energy of triplet excitons is converting triplets to singlets by reverse intersystem crossing (**RISC**) under the assisting of the environment thermal energy, leading to the thermally activated delayed fluorescence (**TADF**).[6] To gain efficient TADF, a small energy gap between singlet and triplet ($\Delta E_{ST}$) is required.[6a]

Recently, we reported a new kind of OLEDs using electrically neutral π-radicals as emitters, wherein the emission comes from the radiative decay of doublets.[7] Because the transition back to the groud state of excited doublet electrons is spin-allowed, the transition problem of triplet excitons is thus circumvented. However, the reported light-emitting neutral π-radicals are all red or yellow emitters, there are no blue light-emitting neutral π-radicals so far.[8] According to the European Broadcasting Union (**EBU**) standard, an excellent blue emitter requires an emission with the Commission Internationale de l'Eclairage coordinates (**CIE$_{(x, y)}$**) of (0.15, 0.06). Whereas, high-performance deep-blue emitters with the CIE coordinates meet or beyond the EBU standard are rare.[9]

In this work, we designed and synthesized two donor-acceptor (**D-A**) type molecules, 4-N-[4-(9-phenylcarbazole)]-3,5-bis(4-diphenlamin)phenyl -4*H*-1,2,4-triazole (**TPA-TAZ**) and 4,4'-(9-(4-(1-phenyl-1H-phenanthro[9,10-d]imidazol-2-yl)phenyl)-9H-carbazole-3,6-diyl) bis-(N,N-diphenylaniline) (**TCP**), as the emitters of deep-blue OLEDs. An optimized TPA-TAZ-based OLED achieves an excellent CIE coordinates of (0.158, 0.043) and a maximun external quantum efficiency (**EQE**) of 6.8 %, which is the largest value of the undoped OLEDs with CIE$_{(y)}$< 0.06 (the EBU blue standard) up to now. The singlet formation ratios of TPA-TAZ and TCP based OLEDs are both higher than the simple spin-statistics of 25 %. Comprehensive experiments demonstrated that the triplet-harvesting processes of TTA and TADF are not dominant in these OLEDs. The extra singlets are attributed to the triplet-polaron interaction (**TPI**) induced upconversion from triplet to singlet according to the experimental and theoretical results.

The TPA-TAZ was synthesized by Buchwald-Hartwig reaction in moderate yield. The detailed synthesis and some basic characterization of the material are given in the **Supporting Information SI-1,2**. The photophysical properties of TPA-TAZ are shown in **Figure 1** and **Table S1**. **Figure 1a** shows the molecular structure along with the electronic structure of the highest occupied molecular orbital (**HOMO**) and the lowest unoccupied molecular orbital (**LUMO**) of TPA-TAZ. The contour plots of the orbitals were calculated by the density functional theory (**DFT**) using Gaussian 09 series of programs with the B3LYP hybrid functional and 6-31G(d) basis set.[10] As can be seen, the LUMO is localized on the triazole and the three benzenes surrounding it, while the HOMO is delocalized on the whole molecule except the carbazole-attached benzene ring and carbazole uinit. The E$_{HOMO}$ of -5.28 eV and E$_{LUMO}$ of -2.12 eV were obtained by Cyclic Voltammetry (**CV**) measurements (**Figure 1b**). **Figure 1c** shows the absorption (**Abs**) and photoluminescent (**PL**) spectra of TPA-TAZ in **THF** solution and film state. From the onset of the abs spectra, the HOMO to LUMO band gap is determined to be 3.22 eV (for solution) and 3.16 eV (for film), which is very close to the value obtained by CV measurements (3.16 eV). The PL of TPA-TAZ film exhibits deep-blue emission peaking at 428 nm with a narrow **FWHM** (full width at half maximum) of 55 nm, revealing its potential of deep-blue emitter for OLEDs. Fluorescence quantum yield ($\phi_{PL}$) of TPA-TAZ film was measured to be 75±2 % in an integrating sphere, which is a high value among the deep-blue emission materials.

To evaluate the electroluminescence properties of TPA-TAZ, OLEDs were fabricated and the structure was optimized to be: ITO / MoO$_3$ (6 nm) / NPB (30 nm) / TPA-TAZ (20 nm) / TPBi (50 nm) / LiF (0.8 nm) / Al (100 nm). Here NPB (N,N′-di-1-naphthyl-N,N′-diphenylbenzidine) is the hole transporting layer, and TPBi (1,3,5-tri(phenyl-2- benzimidazolyl)-benzene) acts as the electron transporting layer. The MoO$_3$ and LiF are used to modify the work function of the electrodes for improving the charge injection. The energy diagram of the materials used in the OLED is shown in Figure 2a. As can be seen from inset of Figure 2b, the EL of the OLED locates in the deep-blue region with a CIE$_{(x,y)}$ of (0.158, 0.043). The EL spectra scarcely change as driving voltages range from 4 V to 7 V (Figure S4). Figure 2c shows the current density (J) – voltage (V) – luminance (L) characteristics of the OLED. A maximum luminance of 7323 cd m$^{-2}$ is obtained at the voltage of 9 V. The maximum EQE is up to be 6.8 %, which is the highest value of the undoped OLEDs with CIE(y) < 0.06 (the EBU blue standard), to the best of our knowledge. Noteworthy, the device exhibits small efficiency roll-off, i.e., the EQE remains 79 % of the maximum value at the luminance of 1000 cd m$^{-2}$ and even 66 % at the maximum luminance of 7323 cd m$^{-2}$. For the device performance of OLEDs with other structures during the device optimization, please see the supplementary information (Figure S5-7 and Table S2). We found the device performance is sensitive to the thickness of TPA-TAZ layer.

For a fluorescent OLED, the EQE can be expressed by the following equation[9b]:

$$\Phi = \chi_S \phi_{PL} \eta_r \eta_{out} \tag{1}$$

wherein $\Phi$ is the EQE, $\chi_S$ is the ratio of singlets to the total excitons, $\phi_{PL}$ is the PL quantum yield of the emitter, $\eta_r$ is the fraction of injected charge carriers that form excitons and can be regarded to be unit in our optimized devices, $\eta_{out}$ is the light out-coupling efficiency which is commonly considered to be 20 ~ 30 %. Using the singlet ratio $\chi_S$ of 25 % and $\phi_{PL}$ of 75±2 % of TPA-TAZ film, the EQE is calculated to be 3.8±0.1 % ~ 5.6±0.2 %, which is lower than the measured value of 6.3±0.6 %. This means there exists extra singlets exceeding the simple spin-statistics of 25 % in our devices. In the first glance, there may be a portion of singlets converted from triplets via TADF, TTA, and etc. In the following sections, we carried out experiments to check these processes.

TADF is a highly efficient tactics for harvesting triplets. A prerequisite of TADF is the small $\Delta E_{ST}$, which commonly should be smaller than 0.3 eV. **Figure 3a** shows the fluorescence and phosphorscence spectra of TPA-TAZ in THF solution at 77K. The first peaks of the two spectra are at 400 nm and 477 nm, respectively, resulting in a $\Delta E_{ST}$ of 0.5 eV. The relatively large $\Delta E_{ST}$ reveals the low possibility of the occurrence of TADF. We performed the transient PL decay measurements using the degassed THF solution. The transient PL exhibits a single exponential decay with the exciton lifetime of 1.35 ns, as shown in **Figure 3b**. There is no delayed fluorescence even the delay time was lengthened to be 1000 ns and the PL intensity declined by 5 orders of magnitude. This indicates that the high singlet ratio of our TPA-TAZ based OLEDs was not benefited from TADF.

TTA is another way for harvesting the triplet. The TTA-induced delayed fluorescence can be simply expressed as: $F=K_{TTA}[T]^2$, here $k_{TTA}$ is the rate parameter of TTA, $[T]$ is the population density of triplet.[5f-5g] $F$ is proportional to the square of the triplet population density. Hence, the TTA-induced delayed fluorescence would be greatly enhanced if the triplet density increases. We carried out experiments to check the occurence of TTA through:

i) *the PL intensity vs the excitation intensity*. The PL intensity should be linearly proportional to the excitation intensity if there is no TTA. However, if TTA occurs, the PL intensity would be non-linearly proportional to the excitation intensity. **Figure 3c** shows the PL intensity of the TPA-TAZ film versus the excitation intensity of a pulse laser at 355 nm. The PL intensity is linearly increased with the increase of the excitation intensity revealing the lack of the TTA. ii) *transient PL decay under low temperature*. The transient PL decay at low temperature could be used to verify the occurrence of TTA by detecting the delayed fluorescence. As **Figure 4b** shows, there is no delayed fluorescence in the transient PL decay measured at 77K. From above experiments, we can confirm that the TTA process is not dominant in TPA-TAZ based OLEDs.

Ma et al. reported a series of D-A type molecules with singlet formation ratio higher than 25 %.[11] They attributed the high singlet ratio to the RISC occuring at the higher-level excited states, e.g., the RISC from $T_2$ to the singlet manifold. A prerequisite of efficient this type RISC is the large energy gap between $T_2$ and $T_1$ (> 1 eV), leading to the inefficient interconvertion from $T_2$ to $T_1$ and thus $T_2$ is converted to singlet. However, time-dependent DFT calculation shows that the energy gap between $T_2$ and $T_1$ of TPA-TAZ is small (just 0.16 eV) (**Figure S8**), which suggests that the high singlet formation ratio of TPA-TAZ based OLEDs does not originate from this process.

Besides TPA-TAZ, we also designed and synthesised another D-A type molecule, TCP, for deep-blue OLEDs. Using TCP as the emitter, a nondoped deep-blue OLED with an excellent CIE coordinates of (0.156, 0.058) was obtained. The OLED has a very simple optimized structure of ITO / $MoO_3$ (5 nm) / TCP (55 nm) / TPBI (55 nm) / LiF (0.5 nm) / Al (100 nm). The singlet ratio is also calculated to be higher than the simple spin-statistics of 25 % according to the measured data of EQE and PL quantum yield of TCP film, and the extra singlets are verified not to benefit from the TTA, TADF, and higher-level RISC processes. The efficiency of the device also displays a slow roll-off of 26 % at the maximum luminance. The details of synthesis and characterization of TCP and TCP based OLEDs are given in the **Supporting Information SII**.

As it has been demonstrated that the TTA, TADF, and higher-level RISC cannot explain why the singlet formation ratio is higher than 25 % expected by the simple spin-statistics in our TPA-TAZ and TCP based devices, the question is what process cause the creation of the extra singlets. It has been theoretically and experimentally investigated that the singlet formation cross section can be larger than that of triplets in some polymers and oligomers,[12] and triplet excitons can be converted to singlet gound state or singlet excitons through TPI process.[13]

The *magneto-current* (**MC**) describes the current change of OLEDs under the application of an external magnetic field (B), and can be expressed as *MC= (Current (B)-Current (0))/ Current (0)*.[14] The TPI process has its own fingerprint MC profile,[15] thus the MC can be used as a feasible tool to verify the TPI process. We proformed the MC measurements of TPA-TAZ and TCP based devices, and the reults are shown in **Figure 4** and **Figure S18**, respectively. We can see that all the experimental data of MC vs B can be well fitted by the formula of $MC= \alpha B^2/(B+B_L)^2+ \beta B^2/(B+B_H)^2$, the first term is the low field effect (**LFE**) and the second term is the high field effect (**HFE**), the $B_L$ and $B_H$ are setted as 5 mT and 100 mT, respectively. The fitted results suggest that there really exists TPI process in our devices.[15]

Both TPA-TAZ and TCP are D-A type molecules, there exists strong couplings between the electron-donor and electron-acceptor moieties of neighbour molecules, which offers the possiblity of converting triplet excitons to singlet excitons through TPI. Different from the two-electron exchange mechanism of TPI proposed in the previous reported works,[13c-13d] here we suggested the one-electron charge transfer mechanism of TPI. There are two possible routes, one is triplet-positive polaron ($P^+$) interaction, the other is triplet-negative porlaron ($P^-$) interaction, as shown in **Figure 5a** and **5b**, respectively. For the triplet-$P^+$ interaction, the electron can hop from the electon-acceptor moietiy of triplet molecule to the electon-acceptor moietiy of its neighbour $P^+$ molecule. If the spin direction of this electron is opposite to that of the hole at the electron-donor moietiy of $P^+$ molecule, the siglet exciton can be created. The similar process can happen for the triplet-$P^-$ interaction, wherein the hole hops between the adjacent molecules. It should be noted that the spin-flip does not need for the TPI induced upconversion from triplet to singlet, which is different from TADF and higher-level RISC.

We deduced the TPI rate through the semi-classical Marcus theory and performed the quantum chemistry computation. The details are given in the **Supporting Information SIII**. According to the deduced electronic coupling formalism (the appendix in SIII-1), we found the coupling of two-electron exchange mechanism is quite small compared to the coulpling of one-electron hopping mechanism of TPI. Therefore, we think the proposed one-electron hopping mechanism in this work is more significant. **Figure 6** shows the six possible charge hopping routes of TPI of TPA-TAZ molecule in crystal. The details of crystal growth and crystal data can be found in the **Supporting Information SI-6**. **Table S6** gives the theoretical prediction of hole (electron) hopping rate $k_{TP^-I}$ ($k_{TP^+I}$) during the triplet-$P^-$ (triplet-$P^+$) interaction process at room temperature. The $k_{TP^-I}$ and $k_{TP^+I}$ achieve as high as $2.00\times10^5$ and $3.79\times10^5$, respectively, which can efficiently compete with the nonradiative transition back to the ground state of triplets. It should be noted that the TPI processes can occur circularly thus continuously convert triplet to singlet, inducing the much higher singlet ratio in OLEDs.

We notice two recently reported D-A molecules, PPI-PPITPA and PPI-PPIPCz.[16] No delayed fluorescence can be observed, but the singlet ratios of the deep-blue OLEDs using them as the emitters are much higher that 25 %. We conjecture there exists the TPI induced conversion from triplet to singlet in their OLEDs. We found the efficiency roll-offs in our TPA-TAZ, TCP based and their PPI-PPITPA, PPI-PPIPCz based OLEDs are not serious, which may be one advantage of the TPI induced upconversion from triplet to singlet.

In summary, we have designed and synthesized two D-A type fluorescent molecules, TPA-TAZ and TCP. Using them as emitters, highly efficient undoped deep blue OLEDs can be obtained. In particular, a TPA-TAZ based OLED achieves an excellent CIE coordinates of (0.158, 0.043), and the maximum EQE of this OLED is up to 6.8 %, which is the largest value of the undoped OLEDs with $CIE_{(y)}$< 0.06 (the EBU blue standard) up to date. The singlet formation ratios of the TPA-TAZ and TCP base OLEDs are both higher than the simple spin-statistics of 25 %. Systematical experiments demonstrated that the triplet-harvesting processes of TTA, TADF and higher-level RISC are not primary origin in the TPA-TAZ and TCP base OLEDs. Instead the TPI-induced upconversion from triplet to singlet through one-electron transfer mechanism is proposed and proven by the magneto-current measurement and quantum chemistry computation. The TPI process does not need the spin-flip, and the efficiency roll-offs of the OLEDs based on TPI are not remarkable. Our results may offer a new way to break through the 25 % upper limit of IQE of fluorescent OLEDs, especially, the deep-blue fluorescent OLEDs.

**Experimental Section**

The Mass spectra and NMR spectra were measured with a Thermo Fisher ITQ1100 GC-MS spectrometer and Bruker AVANCEIII500 spectrometer, respectively. Thermal gravimetric analysis (TGA) was carried on the Pyris1 TGA thermal analysis system at heating rate of 20℃/min in a nitrogen atmosphere. Differential scanning calorimetry (DSC) was recorded on Netzsch DSC204 instrument at a heating rate of 10 $^{o}$C/ min from 20 to 400 $^{o}$C in a nitrogen atmosphere. The CV measurements were performed using an electrochemical analyzer (CHI660C, CH Instruments, USA).

For the Abs and PL measurements, the spectra were measured using a UV-Vis spectrophotometer (Shimadzu UV-2550) and a spectrofluorophotometer (Shimadzu RF-5301PC), respectively. For the fluorescence decay measurements, an Edinburgh fluorescence spectrometer (FLS980) was used. The lifetime of the excited states was measured by time-correlated single photon counting method (detected at the peak of the PL) under the excitation of a laser (375 nm) with the pulse width of 50 ps.

The OLEDs were fabricated by the multiple source organic molecular beam deposition method at the vacuum of $2\times10^{-4}$ Pa. The current density-voltage (J-V) characteristics were measured by a Keithley 2400 source meter. The luminance–voltage (L-V) characteristic and the EL spectrum were measured by a PR650 spectroradiometer. Immediately after fabrication, the OLEDs were placed on a Teflon stage between the poles of an electromagnet with the magnetic field perpendicular to the current. A Keithley2612 sourcemeter was used to provide a constant voltage from channel A, and channel A also recorded the current intensity. The measurements were carried out at room temperature under ambient condition.

CCDC 1419970 contains the supplementary crystallographic data for this paper. These data can be obtained free of charge from The Cambridge Crystallographic Data Centre via www.ccdc.cam.ac.uk/data_request/cif.

**Supporting Information**

Supporting Information is attached in this file.

**Acknowledgements**

We are grateful for financial support from the Ministry of Science and Technology of China (grant number 2015CB655003 and 2013CB933503) , National Natural Science Foundation of China (grant numbers 61275036, 21221063, 21290190 and 91233113) and Graduate Innovation Fund of Jilin University (Project No. 2015028).

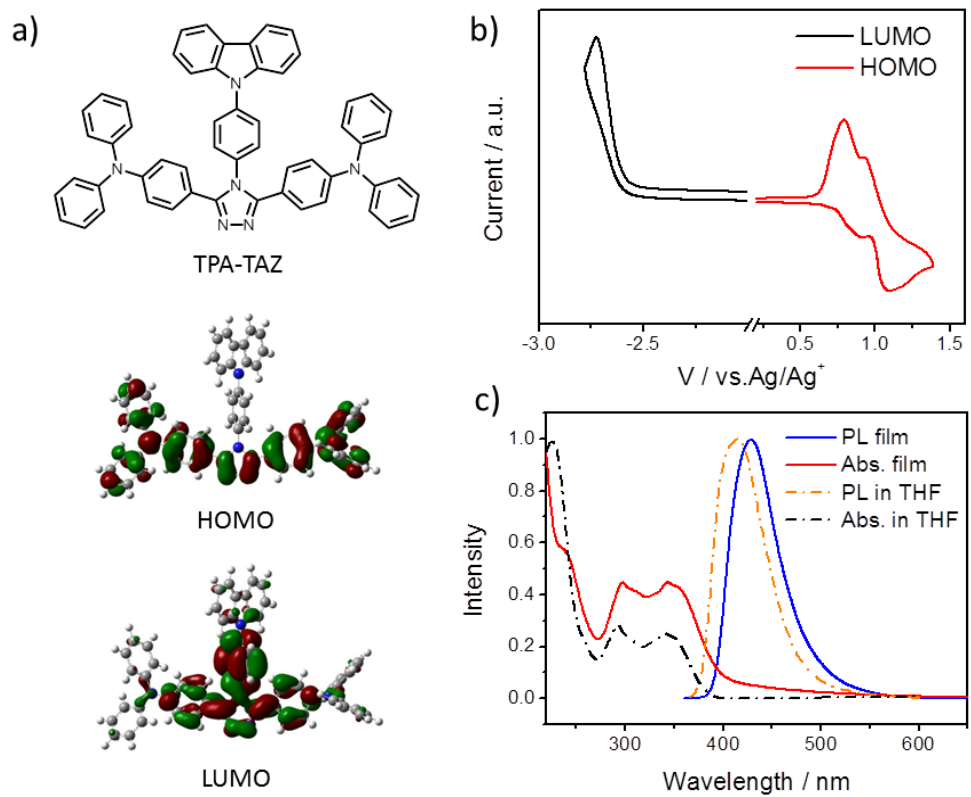

**Figure 1.** a) The molecular structure and electronic structures of HOMO and LUMO of TPA-TAZ; b) The Cyclic Voltammetry spectrum of TPA-TAZ; c) The Abs and PL spectra of the molecule in THF solution and in film state.

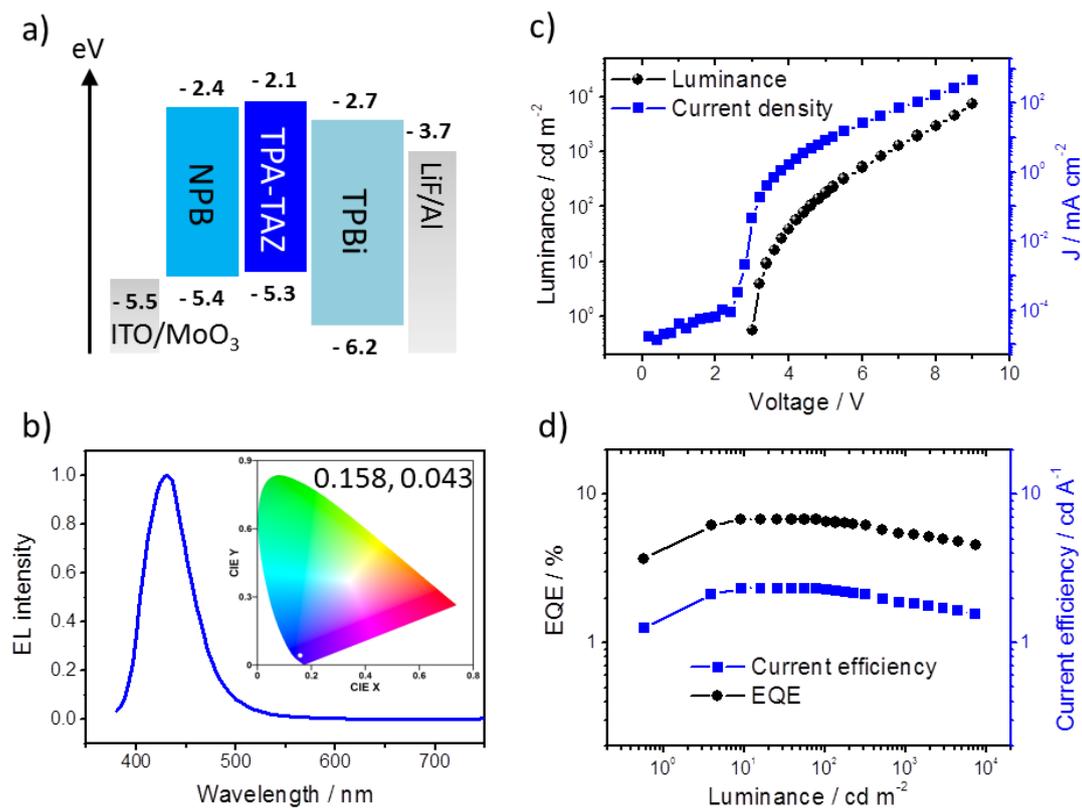

**Figure 2.** a) The schematic diagram of the structure of OLEDs and the energy levels of the materials; b) EL spectrum of the OLED, the inset shows the CIE coordinates of the EL; c) The J-V-L characteristics of the OLED; d) The current efficiency and external quantum efficiency of the OLED as a function of the luminance.

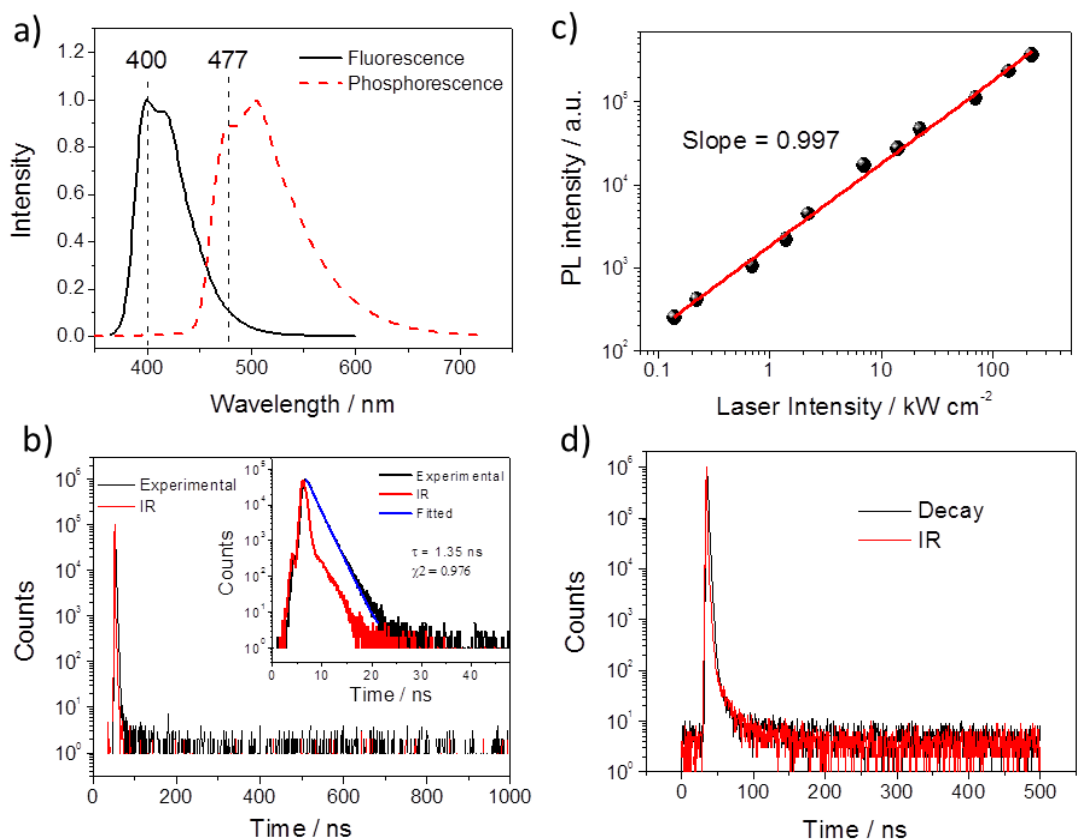

**Figure 3.** a) The fluorescence and phosphorescence spectra of TPA-TAZ in THF solution measured at 77 K; b) The transient PL decay of TPA-TAZ in degassed THF solution tested in the range of 1000 ns, the inset shows the transient PL decay of the same sample test in the range of 50 ns; c) The PL intensity of the TPA-TAZ film as a function of the intensity of the excitation laser; d) The transient PL decay of the TPA-TAZ in THF solution at a temperature of 77 K.

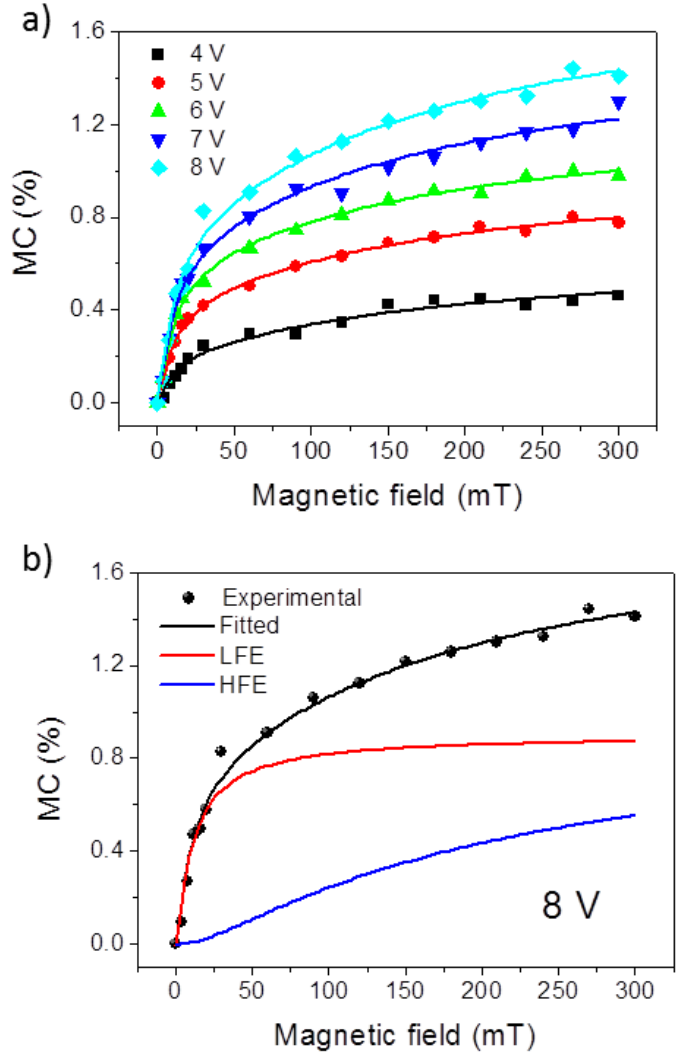

**Figure 4.** a) The MC Vs B at differ driving voltages of TPA-TAZ based OLED, all the MC can be fitted by the formula of $MC = \alpha B^2/(B+B_L)^2 + \beta B^2/(B+B_H)^2$, the first term is the low field effect (LFE) and the second term is the high field effect (HFE); b) The experimental, fitted, LFE and HFE curves at 8 V, the fitted parameters $\alpha$, $\beta$, $B_L$ and $B_H$ are 0.90, 0.99, 5 mT and 100 mT, respectively.

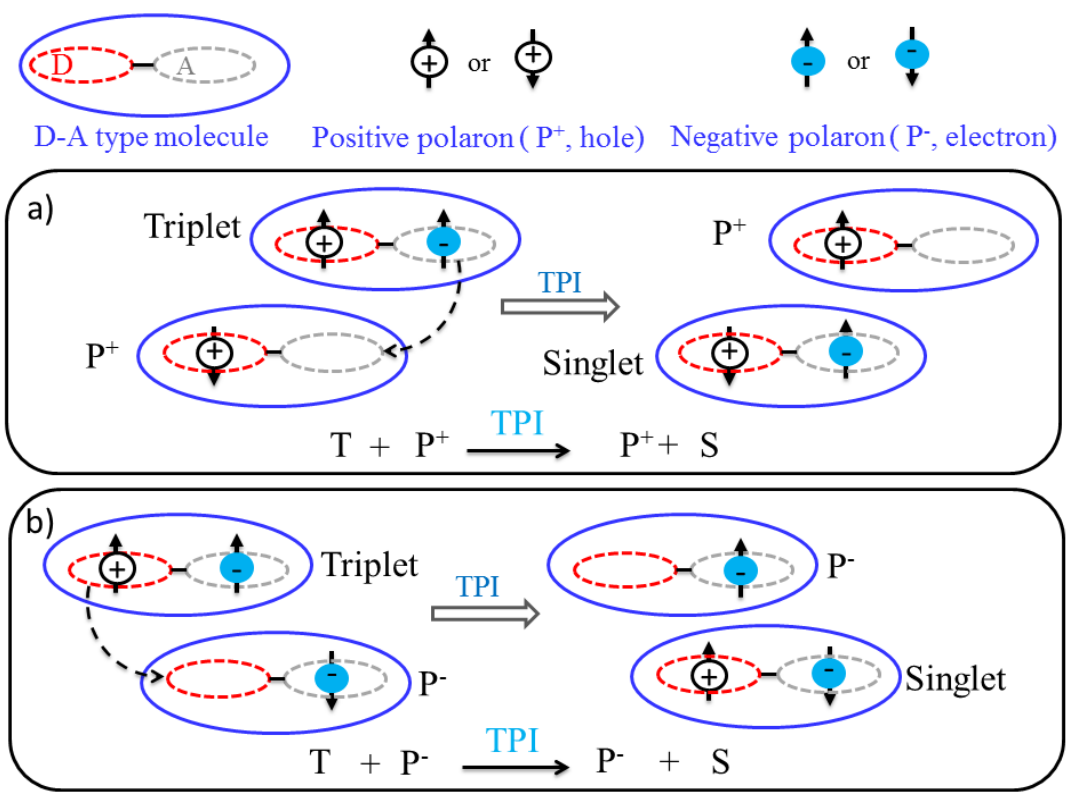

**Figure 5.** Schematic diagram for the TPI induced conversion from triplet to singlet. a) the TPI between triplet and positive-charge polaron; b) the TPI between triplet and negative-charge polaron.

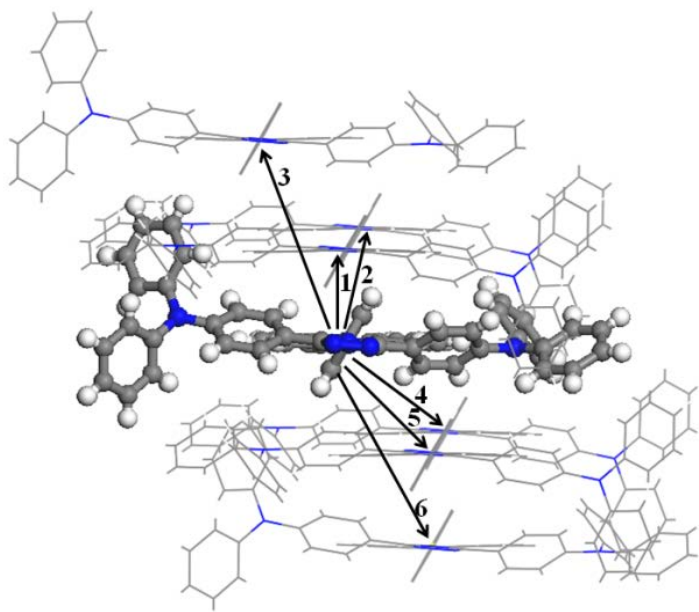

**Figure 6.** Charge hopping routes of TPI of TPA-TAZ molecule in crystal.

**The table of contents:**

Two D-A type molecules, TPA-TAZ and TCP, were designed and synthesized. The OLEDs based on them exhibit deep-blue emission and the singlet formation ratios are higher than the simple spin-statistics of 25 %. The triplet-polaron interaction induced upconversion from triplet to singlet through one-electron transfer mechanism is proposed, and proven by the magneto-current measurement and quantum chemistry computation.

**Keywords**: organic light-emitting device; deep-blue emission; fluorescence; triplet-polaron interacion; charge transfer

Ablikim Obolda, [†] Qiming Peng ,[†] Chuanyou He, [†] Tian Zhang,  Jiajun Ren, Hongwei Ma, Zhigang Shuai*and Feng Li*

**Title:** Triplet-Polaron Interaction Induced Upconversion from Triplet to Singlet: a New Way to Obtain Highly Efficient OLEDs

# Supporting Information

**Triplet-Polaron Interaction Induced Upconversion from Triplet to Singlet: a New Way to Obtain Highly Efficient OLEDs**

*Ablikim Obolda,[†] Qiming Peng,[†] Chuanyou He,[†] Tian Zhang, Jiajun Ren, Hongwei Ma, Zhigang Shuai\* and Feng Li\**

Prof. Dr. Feng Li, Dr. Ablikim Obolda, Dr. Qiming Peng, MS. Chuanyou He, Dr. Hongwei Ma
State Key Laboratory of Supramolecular Structure and Materials, Institute of Theoretical Chemistry, Jilin University, Qianjin Avenue, Changchun, 130012, P. R. China
E-mail: *lifeng01@jlu.edu.cn*
Prof. Dr. Zhigang Shuai, Dr. Tian Zhang, Dr. Jiajun Ren
Key Laboratory of Organic OptoElectronics and Molecular Engineering, Department of Chemistry, Tsinghua University, Beijing 100084, P. R. China
E-mail: *zgshuai@tsinghua.edu.cn*
[†]     these authors contributed equally to this work.

**Contents**

SI-1. Synthsis of TPA-TAZ

SI-2. Thermal and photophysical properties of TPA-TAZ

SI-3. EL spectra of the optimized TPA-TAZ based OLED at different voltages

SI-4. Device performance of TPA-TAZ based OLEDs with other structures

SI-5. DFT calculation of energy levels of TPA-TAZ

SI-6. Details of crystal growth and crystal data of TPA-TAZ

SII-1. Synthesis of TCP

SII-2. Thermal Properties of TCP

SII-3. DFT calculation of TCP

SII-4. Electrochemical properties of TCP

SII-5. Photophysical properties of TCP

SII-6. Electroluminescence properties of TCP

SII-7. The experiments to explore the higher singlet ratio than 25% in TCP-based OLED

SII-8. The MC measurement of TCP based OLED

SIII-1 The methods to calculate the TPI rate

SIII-2 The computational details

SIII-3 The resluts of quantum chemistry computation

**SI-1. Synthsis of TPA-TAZ**
**General:**

All chemicals and solvents, unless otherwise stated, were purchased from commercial suppliers and used as received. THF were distilled before use. Column chromatography was performed using silica gel (200-300 mesh).

The $^1$H-NMR spectra was recorded with a BrukerAVANCEIII500 spectrometer at 500 MHz, using deuterated dichloromethane ($CD_2Cl_2$) as the solvent at 298 K. Ultraviolet-visible (UV-vis) absorption spectra was recorded on a Shimadzu UV-2550 spectrophotometer. Fluorescence spectra was performed using a RF-5301PC spectrophotometer. Mass spectra of all compounds were measured on Thermo Fisher ITQ1100 mass detector.

**Synthesis:**

The synthesis routes of the compounds are given in the Scheme S1. All of the intermediates, 1-5, were synthesized according to the procedures reported previously.[1,2] The last target molecule 4-N-[4-(9-phenylcarbazole)]-3,5-bis(4-diphenlamin) phenyl-4*H*-1,2,4-triazole (TPA-TAZ) was synthesized by Buchwald–Hartwig reaction, using of intermediate5 and diphenylamine in the presence of palladium acetate, sodium tert-butoxide and tri-tert-butylphosphine.

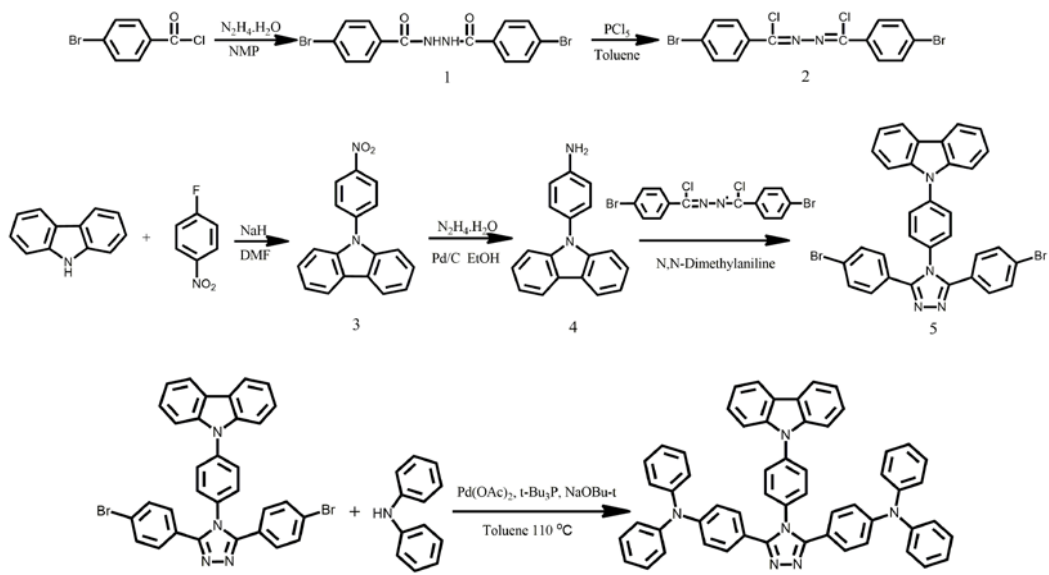

**Scheme S1.** Synthetic routes of the intermediates and TPA-TAZ

Under the nitrogen atmosphere, palladium acetate (65 mg, 0.097 mmol), tri-t-butylphosphine (58 mg, 0.29 mmol) and toluene were stirred at room temperature about 15 minutes to obtain a palladium-phosphine complex. Then 4-N-[(4-(9-phenylcarbazole)]-3,5-bis(4-bromophenyl) -4$H$-1,2,4-triazole (**5**) (1.0 g, 1.61 mmol), sodium tert-butoxide ( 0.464 g, 4.83 mmol), diphenylamine (0.817 g, 4.83mmol) were added with 50 mL toluene and reaction mixture heated up to 110°C and stirred for 36 h to complete the reaction. After cooling to room temperature, 80 mL water was added and extracted with $CH_2Cl_2$ several times, washed with water, organic layer was combined and dried over anhydrous magnesium sulphate. After the solvent was evaporated under vacuum, the crude product directly absorbed in silica gel and purified by $SiO_2$ column chromatography twice (ethyl acetate: dichloromethane = 1:5), solvent was evaporated and dried in vacuum to afford product TPA-TAZ (0.72 g, 56 % yield), to obtain more pure product TPA-TAZ recrystallized with dichloromethane and ethanol. $^1$H NMR (500 MHz, $CD_2Cl_2$) δ [ppm] = 8.20 (d, J = 7.6 Hz, 2H), 7.72 (d, J = 8.3 Hz, 2H), 7.51 (d, J = 8.3 Hz, 2H), 7.48 – 7.36 (m, 10H), 7.33 (t, J = 7.8 Hz, 8H), 7.17 (d, J = 7.8 Hz, 8H), 7.13 (t, J = 7.3 Hz, 4H), 7.04 (d, J = 8.6 Hz, 4H).$^{13}$C NMR (126 MHz, $CD_2Cl_2$) δ[ppm] =154.35, 149.18, 147.02, 140.39, 138.65,

134.28, 129.69, 129.50, 129.45, 128.14, 126.20, 125.32, 123.89, 123.61, 121.34, 120.48, 120.38, 119.73, 109.46. MS (m/z): calcd for $C_{56}H_{40}N_6$, 796.33; found, 796.52[M+]. Elem. Anal. Calcd for $C_{56}H_{40}N_6$: C, 84.40; H, 5.06; N, 10.55. Found: C, 84.38; H, 4.978; N, 10.63.

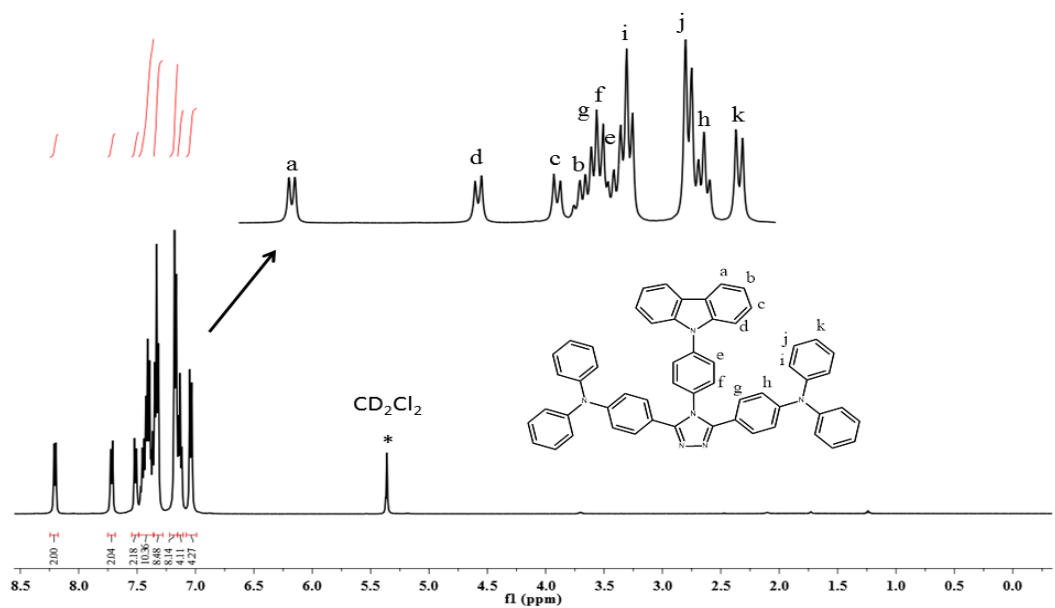

**Figure S1.** $^1$H-NMR Spectrum of TPA-TAZ in $CD_2Cl_2$

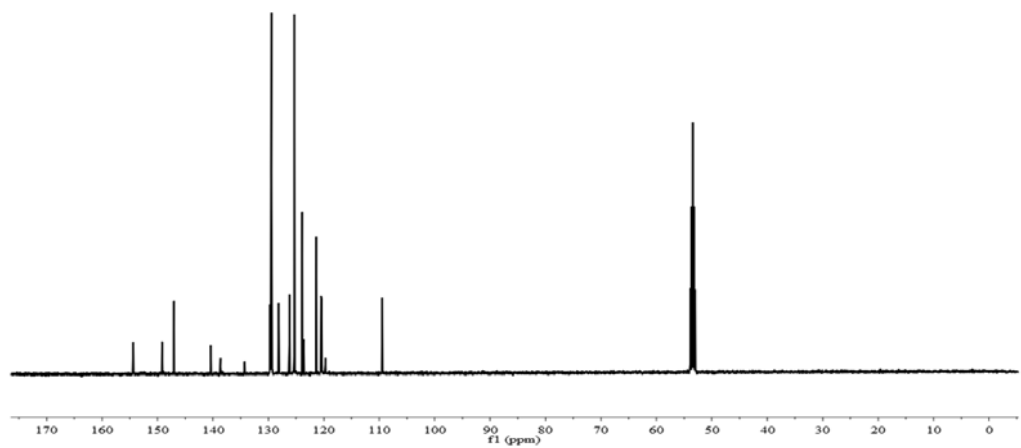

**Figure S2.** $^{13}$C-NMR Spectrum of TPA-TAZ in $CD_2Cl_2$

## SI-2. Thermal and photophysical properties of TPA-TAZ

Thermal gravimetric analysis (TGA) was carried on the Pyris1 TGA thermal analysis system at heating rate of 20 °C min$^{-1}$ under nitrogen protection. A decomposition temperature of 480 °C was obtained.

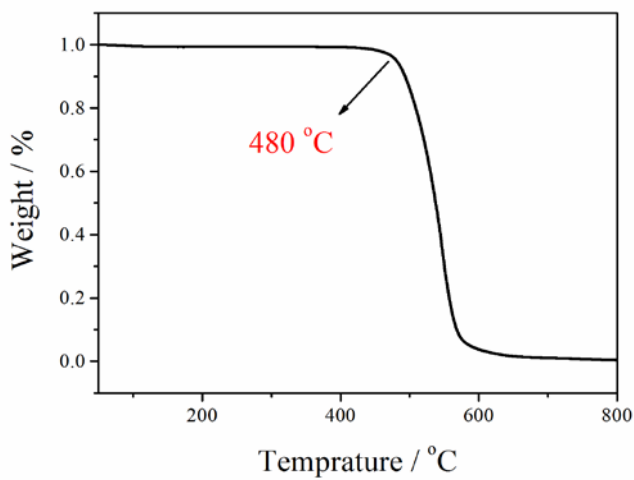

**Figure S3.** TGA curve of TPA-TAZ under nitrogen flow.

**Table S1**. Photophysical properties of TPA-TAZ.

|  | $\lambda_{abs}$[a)] [nm] | $\lambda_{PL}$[b)] [nm] | HOMO[c)] [eV] | Eg [eV] | LUMO[c)] [eV] |
| --- | --- | --- | --- | --- | --- |
| In THF | 339 (385) | 415 (377) | -5.28 | 3.16 | -2.12 |
| Film | 341 (393) | 428 (390) |  |  |  |

[a)] The first peak of the absorption spectrum (the onset of the absorption spectrum); [b)] The peak of the PL spectrum (the onset of the PL spectrum); [c)] Obtained from the Cyclic Voltammetry measurements.

**SI-3. EL spectra of the optimized TPA-TAZ OLED at different voltages**

As can be seen from Figure S4, the EL spectra of the optimized OLED at different voltages overlap very well.

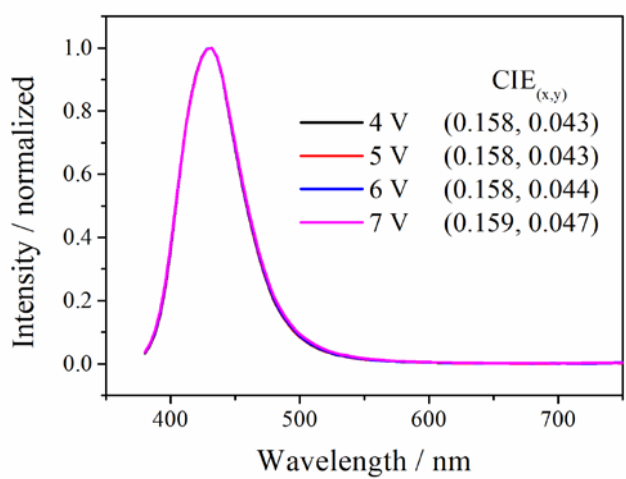

**Figure S4.** The EL spectra of the optimized OLED at different voltages ranging from 4 to 7 V.

**SI-4. Device performance of TPA-TAZ OLEDs with other structures**

**Device I:** ITO/MoO$_3$(6nm)/TPA-TAZ(55 nm)/TPBi (45nm)/LiF(0.8 nm)/Al (100nm);

**Device II:** ITO/MoO$_3$(6 nm)/NPB(35 nm)/TPA-TAZ(25 nm)/TPBi(55 nm)/LiF(0.8 nm)/Al(100 nm)

**Device III:** ITO/MoO$_3$(6 nm)/NPB(35 nm)/TPA-TAZ(20 nm)/TPBi(55 nm)/LiF(0.8 nm)/Al(100 nm)

**Table S2.** Performance of the Devices I, II, and III.

|  | $L_{max}$ (cd m$^{-2}$) | $EL_{peak}$ (nm) | $CE_{max}$ (cd A$^{-1}$) | $EQE_{max}$ (%) | $CIE_{(x,y)}$ |
|---|---|---|---|---|---|
| Device I | 3936 | 428 | 1.80 (4.8 V) | 4.16 (4.8 V) | 0.158, 0.056 |
| Device II | 6843 | 432 | 2.29 (4.0 V) | 6.36 (4.0 V) | 0.158, 0.054 |
| Device III | 7139 | 432 | 2.46 (3.6 V) | 6.08 (3.6 V) | 0.158, 0.051 |

$L_{max}$, the maximum luminance; $EL_{peak}$, the peak of the EL spectrum; $CE_{max}$, the maximum current efficiency of the OLEDs measured at a corresponding voltage; $EQE_{max}$ the maximum external quantum efficiency of the OLED measured at a corresponding voltage; $CIE_{(x,y)}$, the CIE coordinates.

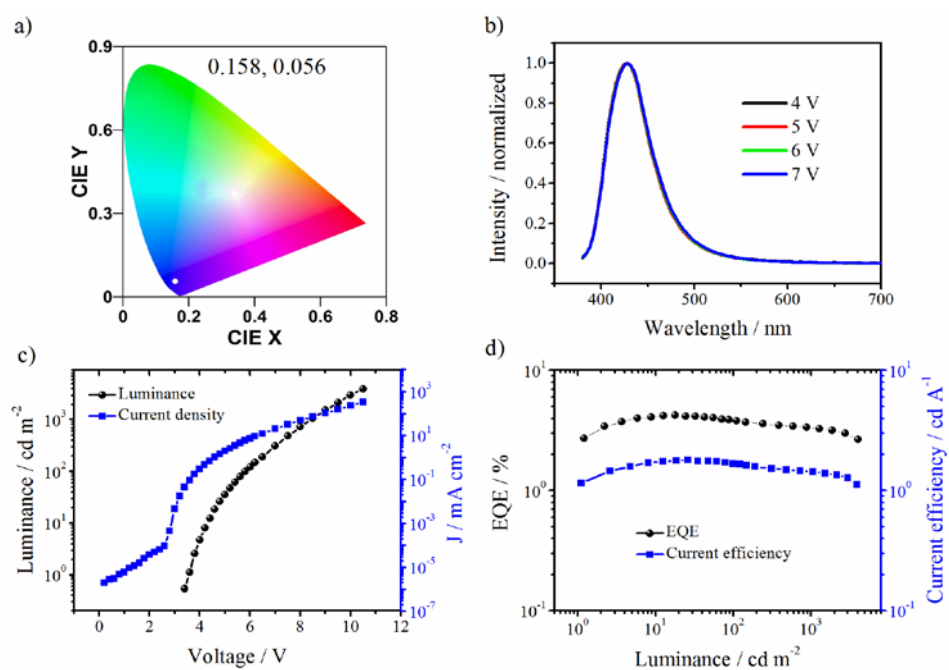

**Figure S5.** Performance of Device I. a) The CIE coordinates; b) The EL spectra; c) the J - V- L characteristics; d) The current efficiency and EQE Vs luminance.

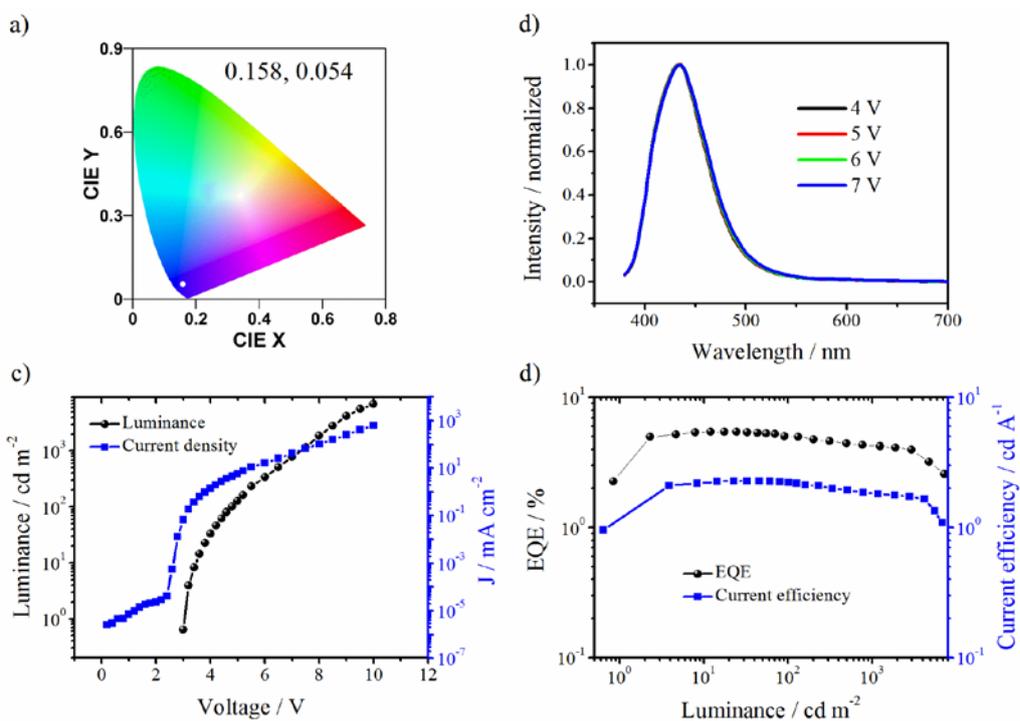

**Figure S6.** Performance of Device II. a) The CIE coordinates; b) The EL spectra; c) the J - V - L characteristics; d) The current efficiency and the EQE Vs luminance.

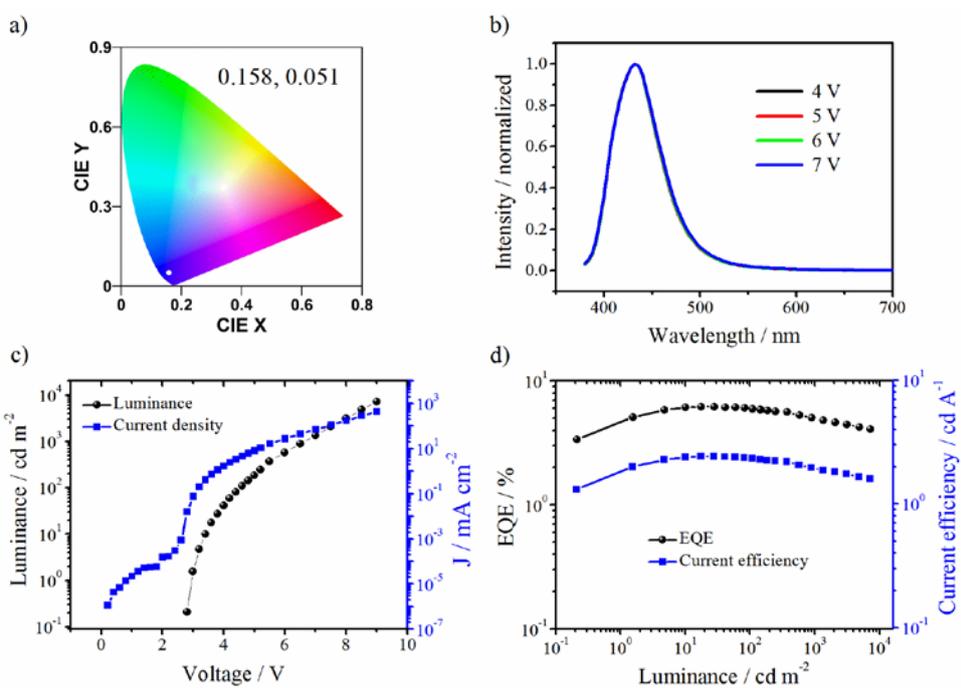

**Figure S7**. Performance of Device III. a) The CIE coordinates; b) The EL spectra; c) the J - V - L characteristics; d) The current efficiency and the EQE Vs luminance.

**SI-5. DFT calculation of energy levels of TPA-TAZ**

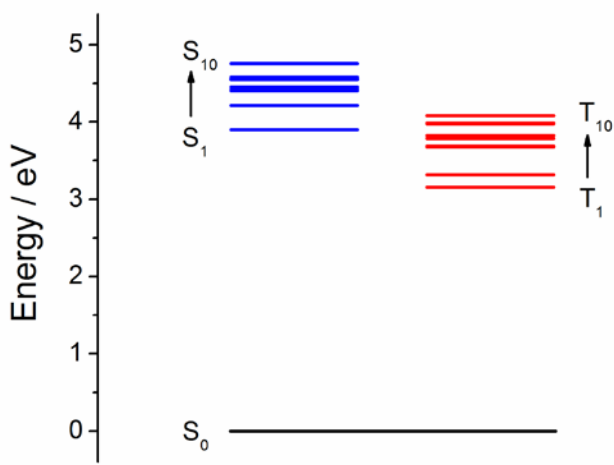

**Figure S8.** The calculated energy levels of S1 to S10 and T1 to T10 of TPA-TAZ from time-dependent DFT calculation.

**SI-6. Details of crystal growth and crystal data of TPA-TAZ**

TAP-TAZ single crystals were successfully obtained by the solution diffusion method wherein dichloromethane was used as the favorable solvent and ethanol as the unfavorable solvent.

Single-crystal X-ray diffraction data were collected using a Rigaku RAXIS-PRID diffractometer with graphite monochromator Mo K$\alpha$ radiation. The structure was solved with direct methods using the SHELXTL programs and refined with full-matrix least squares on $F^2$. Anisotropic thermal parameters were refined for all the non-hydrogen atoms. The positions of hydrogen atoms were located from difference maps and refined isotropically. CCDC 1419970 containing the crystallographic data can be obtained free from the Cambridge Crystallographic Data Centre via www.ccdc.cam.ac.uk/data_request/cif. Crystal data and structure refinement are listed in Table S3.

**Table S3.** Crystal data and structure refinement of TAP-TAZ.

| Identification code | tpataz |
|---|---|
| Empirical formula | $C_{57} H_{43} N_6 O_{0.50}$ |
| Formula weight | 819.97 |
| Temperature | 293(2) K |
| Wavelength | 0.71073 A |
| Crystal system, space group | Triclinic, P -1 |
| Unit cell dimensions | a = 9.6110(19) A    alpha = 99.88(3) deg |
| | b = 13.412(3) A    beta = 96.84(3) deg. |
| | c = 18.755(4) A    gamma = 106.34(3) deg. |
| Volume | 2249.5(8) A^3 |
| Z, Calculated density | 2,   1.211 Mg/m^3 |
| Absorption coefficient | 0.073 mm^-1 |
| F(000) | 862 |
| Crystal size | 0.50 x 0.21 x 0.13 mm |
| Theta range for data collection | 3.03 to 25.00 deg. |
| Limiting indices | -11<=h<=11, -15<=k<=15, -22<=l<=22 |
| Reflections collected / unique | 16620 / 7611 [R(int) = 0.0598] |
| Completeness to theta = 25.00 | 96.1 % |
| Absorption correction | Semi-empirical from equivalents |
| Max. and min. transmission | 0.9906 and 0.9649 |
| Refinement method | Full-matrix least-squares on F^2 |
| Data / restraints / parameters | 7611 / 21 / 589 |
| Goodness-of-fit on F^2 | 0.964 |
| Final R indices [I>2sigma(I)] | R1 = 0.0727, wR2 = 0.1993 |
| R indices (all data) | R1 = 0.1372, wR2 = 0.2600 |
| Extinction coefficient | 0.0054(14) |
| Largest diff. peak and hole | 0.583 and -0.266 e.A^-3 |

## SII-1. Synthesis of TCP

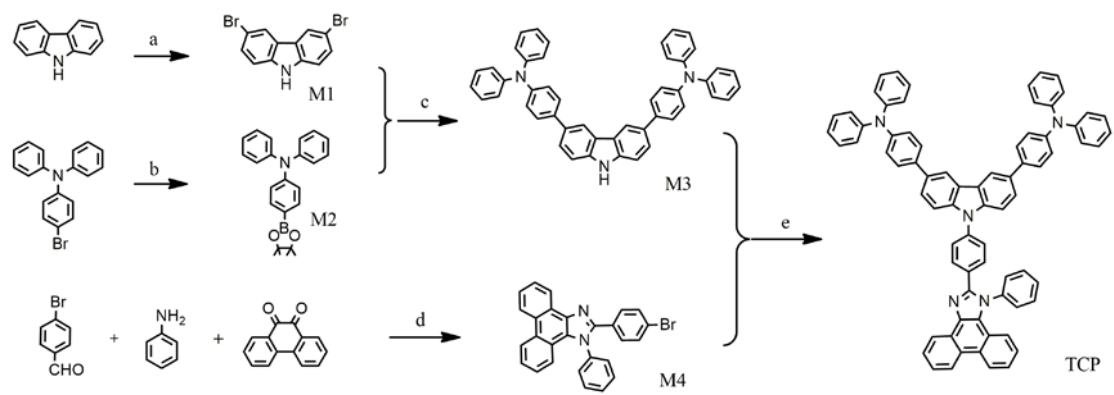

**Scheme S2.** The synthetic route to TCP.

TCP was synthesized by an Ullmann coupling reaction with carbazole substituted with triphenylamine and bromo-imidazole with a yield of 63.8 %. Scheme 1 shows the synthetic route to TCP. All the solvents and materials were gotten from commercial company without further purification.

**SII-1-1. Synthesis of 3,6-dibromo-9-H-carbazole (M1).**

Carbazole (5 g, 30 mmol) was dissolved in $CHCl_3$. Then N-bromosuccinimide (NBS, 11.7 g, 63 mmol) was added in batches away from light, and the reaction solution was continuously stirred for 12 h. The resulting mixture was stirred for another 30 min with appropriate water, and extracted with $CH_2Cl_2$. The $CH_2Cl_2$ solution was dried over anhydrous $MgSO_4$. After removing $CH_2Cl_2$ solvent, the residue was further purified by silica gel column chromatography ($CH_2Cl_2$/petroleum ether (1:1)) to give 8.2 g (84.1 %) of a white powder. $^1$H NMR (500 MHz, DMSO) δ 11.61 (s, 1H), 8.44 (d, J = 1.7 Hz, 2H), 7.54 (dd, J = 8.6, 1.9 Hz, 2H), 7.48 (d, J = 8.6 Hz, 2H). m/z: 325.04 [M$^+$] (calcd: 325.00).

**SII-1-2. Synthesis of N,N-diphenyl-4-(4,4,5,5-tetramethyl-1,3,2-dioxaborolan-2-yl)aniline (M2).**

4-bromo-N,N-diphenylaniline(5g, 15.4mmol), bis(pinacolato)diboron(4.7 g, 18.5 mmol), Pd(dppf)$Cl_2$ (630 mg, 0.77 mmol), and KOAc (9.1 g, 92.4 mmol) were mixed in a 250 mL flask containing 1,4-dioxane(60 mL). The reaction mixture was refluxed for 24h under nitrogen. After it was cooled to room temperature, a dilute hydrochloric acid solution quenched. The mixture was extracted with $CH_2Cl_2$ and dried over anhydrous $MgSO_4$. The crude product was concentrated by rotary evaporation and further purified by silica gel column chromatography ($CH_2Cl_2$ / petroleum ether (1:1,v/v)) to afford a white solid (3.08 g, 61.5 %). $^1$H NMR (500 MHz, DMSO) δ 7.55 (d, J = 8.2 Hz, 2H), 7.34 (t, J = 7.7 Hz, 4H), 7.10 (dd, J = 14.7, 7.4 Hz, 2H), 7.06 (d, J = 8.1 Hz, 4H), 6.90 (d, J = 8.2 Hz, 2H), 1.28 (s, 12H). m/z: 371.16 [M$^+$] (calcd: 371.28).

**SII-1-3 Synthesis of 4,4'-(9H-carbazole-3,6-diyl)bis(N,N-diphenylaniline) (M3).**

N,N-diphenyl-4-(4,4,5,5-tetramethyl-1,3,2-dioxaborolan-2-yl)aniline ( M2, 4.6 g, 15.37 mmol), 3,6-dibromo-9-H-carbazole (M1, 2.0 g, 6.15 mmol), Pd(PPh$_3$)$_4$ (213.7 mg, 0.18 mmol), $K_2CO_3$ (12 mL, 2 M), and ethanol (6 mL) were mixed in a 100 mL flask containing anhydrous toluene (20 mL). The mixture was refluxed for 48 h under nitrogen. After cooled to room temperature, the reaction mixture was quenched with dilute hydrochloric acid solution and extracted with $CH_2Cl_2$. The organic extracts were dried over anhydrous $MgSO_4$ and concentrated by rotary evaporation. The crude product was further purified by silica gel column chromatography ($CH_2Cl_2$/petroleum ether (1:1,v/v)) to get M3 as white solid (2.86 g, 71.2 %). $^1$H NMR (500 MHz, DMSO) δ 11.36 (s, 1H), 8.53 (s, 2H), 7.76 – 7.65 (m, 6H), 7.55

(d, J = 8.4 Hz, 2H), 7.33 (t, J = 7.8 Hz, 8H), 7.14 – 7.00 (m, 16H). m/z: 653.73 [M$^+$] (calcd: 653.81)

**SII-1-4. Synthesis of 2-(4-bromophenyl)-1-phenyl-1H-phenanthro [9,10-d]imidazole (M4).**

A mixture of phenanthrenequinone (3.1 g, 14.9 mmol), aniline (6.8 mL, 50.0 mmol), 4-bromobenzaldehyde (3.3 g, 17.9 mmol), ammonium acetate (4.6 g, 59.6 mmol), and acetic acid (30 mL) was refluxed for 2 h under nitrogen. When the mixture was cooled to room temperature, the precipitation was collected by vacuum filtration. The crude product was further purified by silica gel column chromatography using $CH_2Cl_2$ as an eluent. We got a white powder (4.98 g, 74.6 %). $^1$H NMR (500 MHz, CDCl$_3$) δ 8.89 (d, J = 7.7 Hz, 1H), 8.79 (d, J = 8.3 Hz, 1H), 8.73 (d, J = 8.3 Hz, 1H), 7.78 (t, J = 7.4 Hz, 1H), 7.67 (ddd, J = 19.7, 12.0, 4.0 Hz, 5H), 7.59 – 7.41 (m, 7H), 7.19 (d, J = 8.2 Hz, 1H). m/z: 448.71 [M$^+$] (calcd: 449.34).

**SII-1-5. Synthesis of TCP.**

4,4'-(9H-carbazole-3,6-diyl)bis(N,N-diphenylaniline) (M3, 1.5 g, 2.3 mmol), 2-(4-bromophenyl)-1-phenyl-1H-phenanthro[9,10-d]imidazole (M4, 1.24 g, 2.76 mmol), K$_2$CO$_3$ (634 mg, 4.6 mmol), 18-crown-6 (42.5 mg, 0.16 mol) and CuI (26.2 mg, 0.14 mmol) were added into a 100 mL flask containing 20 mL o-dichlorobenzene. The reaction mixture was refluxed at 180 oC for 48h under nitrogen. When cooled to room temperature, the mixture was washed with dilute hydrochloric acid solution, ammonium acetate saturated solution and extracted with CH$_2$Cl$_2$. The organic extracts were dried over MgSO$_4$ and concentrated. The crude product was further purified by silica gel column chromatography using CH$_2$Cl$_2$/ petroleum ether (4:1, v/v) as an eluent. And a white solid (1.5 g, 63.8 %) was got. $^1$H NMR (500 MHz, DMSO) δ 8.97 (d, J = 8.5 Hz, 1H), 8.92 (d, J = 8.2 Hz, 1H), 8.76 (d, J = 7.8 Hz, 1H), 8.69 (s, 2H), 7.92 (d, J = 8.5 Hz, 2H), 7.88 – 7.69 (m, 16H), 7.61 – 7.57 (m, 1H), 7.48 (d, J = 8.6 Hz, 2H), 7.40 – 7.30 (m, 9H), 7.14 – 7.05 (m, 16H). $^{13}$C NMR (126 MHz, CDCl$_3$) δ 147.84, 146.67, 140.21, 136.06, 133.43, 130.82, 130.39, 129.26, 129.19, 128.41, 127.91, 127.39, 126.39, 125.35, 124.39, 124.27, 124.19, 123.17, 122.77, 120.91, 118.39, 110.17. MS (m/z): calcd for C$_{75}$H$_{51}$N$_5$, 1022.24; found, 1021.67 [M$^+$]. Elem. Anal. Calcd for C$_{75}$H$_{51}$N$_5$: C, 88.12; H, 5.03; N, 6.85. Found: C, 88.13; H, 5.031; N, 6.83.

**SII-2. Thermal Properties of TCP**

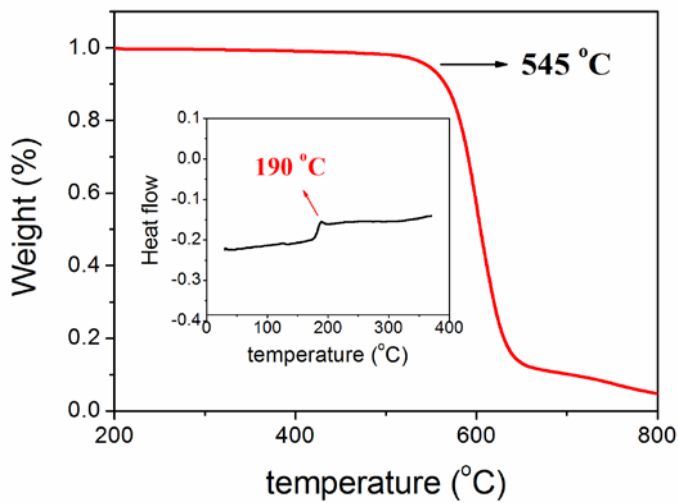

**Figure S9.** Thermogravimetric analysis (TGA) and differential scanning calorimetry (DSC) (the inset) curves of TCP.

Because of the large rigid plane, TCP shows the high thermal stability with a decomposition temperature ($T_d$, corresponding to 5 % weight loss) of 545 °C, as shown in Figure S9. The glass transition temperature ($T_g$) of TCP was measured to be as high as 190 °C (see the inset in Figure S8), which would greatly benefit the stability of the film morphology.

**SII-3. DFT theoretical calculations of TCP**

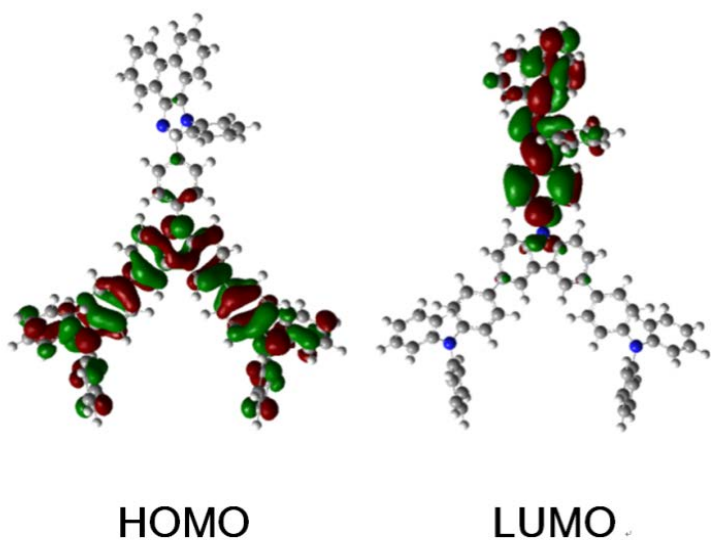

**Figure S10.** The electronic structures of the HOMO and LUMO of TCP.

The optimized molecular geometry and the electronic structures of the HOMO and LUMO of TCP were obtained by density functional theory (DFT) calculations using Gaussian 09 series of programs with the B3LYP hybrid functional and 6-31G(d) basis set, as shown in Figure S10. It can be seen the HOMO is localized on the 3,6-disubstitued carbazole, while the LUMO mainly lies on the phenanthroimidazole group, indicating the ICT character of the molecule.

**SII-4. Electrochemical properties of TCP**

Cyclic Voltammetry (CV) measurements were carried out to get the energy levels of the HOMO and LUMO of TCP. Figure S11 shows the CV curves of TCP. From the onset of the oxidation and reduction curves, and using ferrocene as the reference, the HOMO and LUMO levels were estimated to be -5.14 and -2.14 eV, respectively. The energy gap between HOMO and LUMO ($E_g$) is calculated to be 3.0 eV indicating a deep-blue emission.

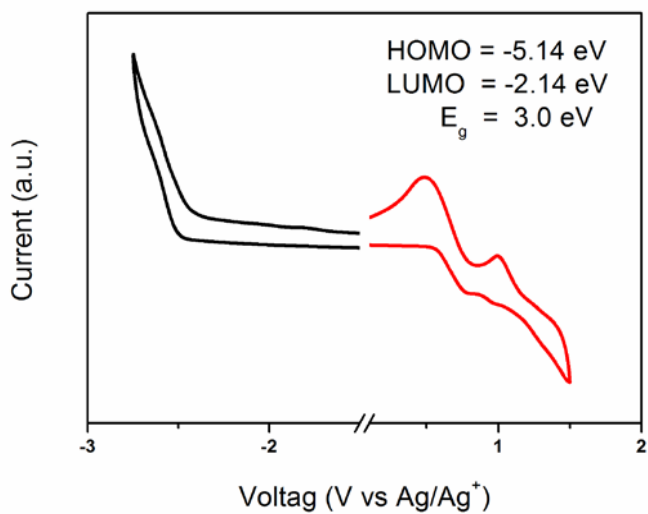

**Figure S11.** Cyclic Voltammograms of TCP.

**SII-5. Photophysical properties of TCP**

In the Abs spectra (Figure 12 a), a peak at around 345 nm can be found, which is attributed to the ICT transition. In the PL spectra, TCP exhibits a single emission peak at around 420 nm in THF solution and 440 nm in film. The PL spectra of TCP lack the fine structure indicating that the emission is originated from the transition of ICT state. To verify this, we checked the solvation effects of TCP. As can be seen from Figure S12 b), the PL spectra show red-shifted, broadened and structureless emissions when increasing the polarity of the solvent. The PL quantum yield ($\phi_{PL}$) of the TCP film was measured to be 0.43 using an integrating sphere system.

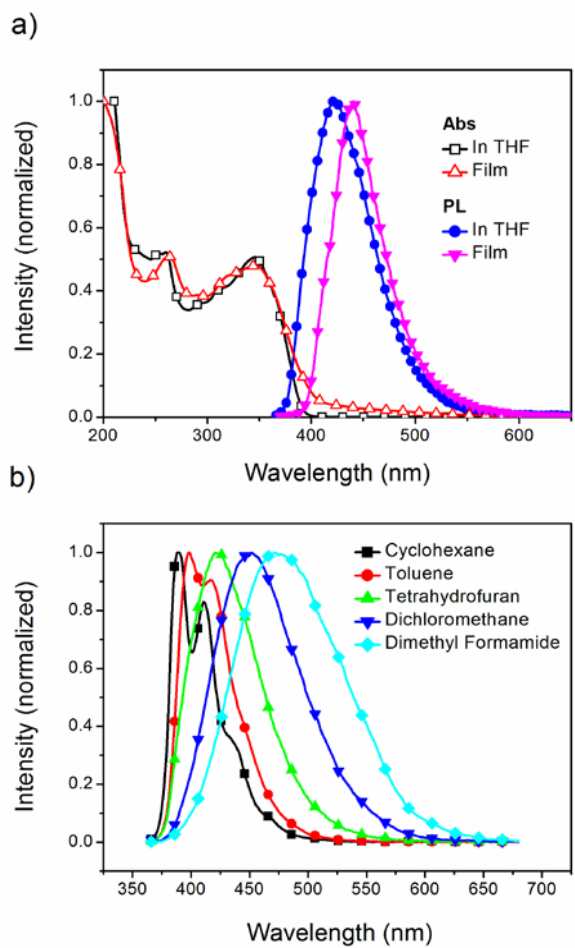

**Figure S12.** a) The Abs and PL spectra of TCP in THF solution($10^{-5}$ mol/L) and film; b) PL spectra of TCP in solvents with different polarity.

**SII-6. Electroluminescence properties of TCP**

We made a deep-blue OLED using TCP as the emitter as well as the the hole-transporting layer with the optimized stucture of ITO / $MoO_3$ (5 nm) / TCP (55 nm) / TPBI (55 nm) / LiF (0.5 nm) / Al (100 nm). The EL spectrum at 4 V is shown in Figure S13 (c) correspdongding to the CIE coordinates of (0.158, 0.058), and the EL spectra barely change over the whole voltages ranging from 3 to 7 V. The maximum luminance ($L_{max}$) and EQE ($EQE_{max}$) are 7715 cd/m$^2$, and 3.5 %, respectively. For an OLED, its EQE can be expressed as: $EQE = \phi_{PL} \times \eta_r \times \eta_{out} \times \chi_S$, wherein, $\phi_{PL}$ is the PL quantum yield of the emitter, $\eta_r$ is the charge balance parameter, $\eta_{out}$ is the light out-coupling efficiency, $\chi_S$ is the singlet ratio to the total excitons. A $\chi_S$ of 27.2 - 40.8 % of this OLED is obtained by using the $\phi_{PL}$ of 0.43 of the TCP film, assuming a $\eta_{out}$ of 20-30 %, and considering the charge balance parameter to be unit. So the singlet ratio is higher than 25 % expected by the simple spin-statistics.

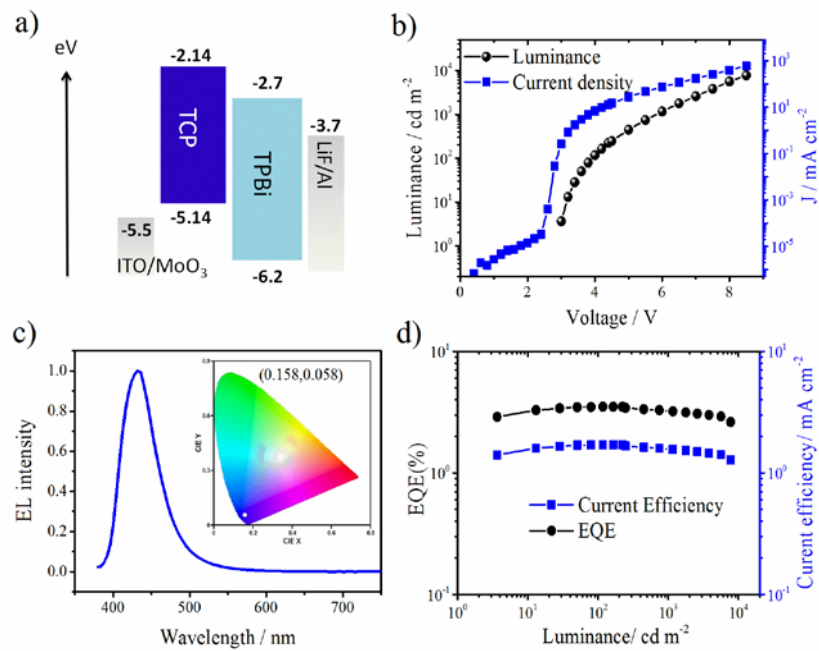

**Figure S13.** a) The schematic diagram of the structure of TCP based OLEDs and the energy levels of the materials; b) The J-V-L characteristics of the OLED; c) EL spectrum of the OLED at 4 V, the inset shows the CIE coordinates; d) The current efficiency and external quantum efficiency of the OLED as a function of the luminance.

**SII-7. The experiments to explore the higher singlet ratio than 25% in TCP based OLED**

**SII-7-1. The fluorescence and phosphorescence spectra of TCP in THF solution**

The fluorescence and phosphorescence spectra of TCP were measured at 77 K in the THF solution with a concentration of $10^{-5}$ mol/L, as shown in Figure S14. An Edinburgh fluorescence spectrometer system (LP920) was used. The energy gap between $S_1$ and $T_1$ ($\Delta E_{ST}$) is calculated to be 0.48 eV according to the peaks of the fluorescence (424 nm) and phosphorescence (507 nm) spectra. The relatively large $\Delta E_{ST}$ suggests the occurrence of the TADF in our TCP-based OLED has the low possibility.

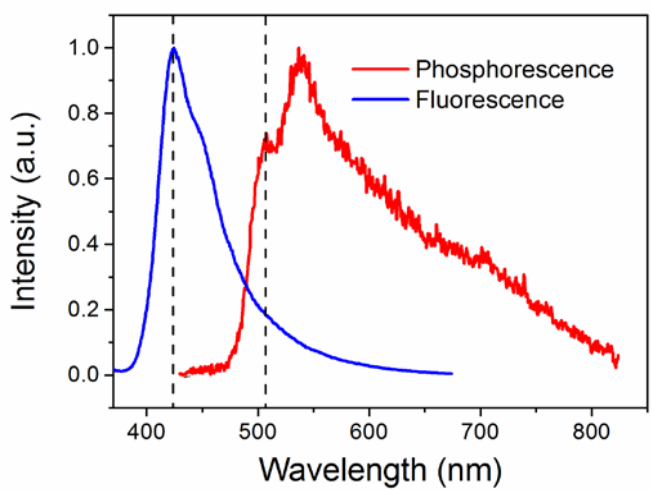

**Figure S14.** The fluorescence and phosphorescence spectra of TCP in THF solution at 77 K.

**SII-7-2. Transient PL decay of TCP in degassed THF solution.**

    The Transient PL decay of TCP in degassed THF solution was also measured at room temperature to test the TADF. As can be seen from Figure S15, the transient PL exhibits a single exponential decay with the exciton lifetime of 2.72 ns. There is no delayed fluorescence can be observed even the delay time was lengthened to be 2000 ns and the PL intensity decreased by 5 orders of magnitude. This result further indicates that the higher singlet ratio of our TCP-based OLED is not benefited from the TADF.

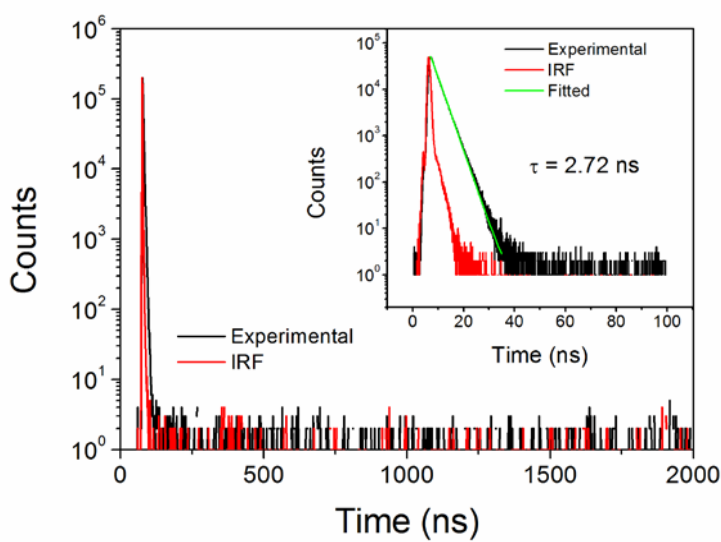

**Figure S15.** The transient PL decay of TCP in degassed THF solution at room temperature tested in the range of 2000 ns, the inset shows the transient PL decay of the same sample test in the range of 100 ns.

## SII-7-3. The experiments to test the TTA process

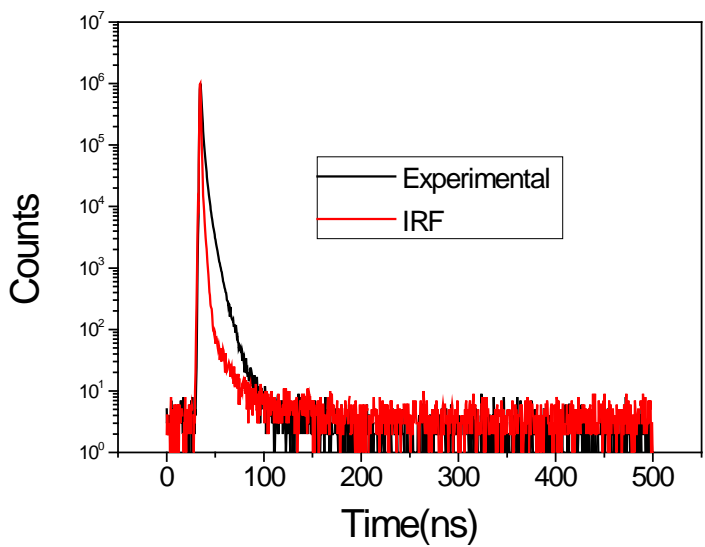

**Figure S16.** The transient PL decay of the TCP in THF solution at a temperature of 77 K.

In the TTA process, one singlet can be generated by consuming two triplets, resulting in the P-type delayed fluorescence (DF). P-type DF is proportional to the square of the triplet population density. Hence, the P-type DF would be greatly enhanced if the triplet population density increases. A feasible method to increase the triplet density is to reduce the temperature and thus suppress the non-radiative decay of the triplets. So we performed the transient PL decay under low temperature to check the occurrence of the TTA. As Figure S16 shows, there is no delayed fluorescence observed, revealing the lack of the TTA.

## SII-7-4. The calculated energy levels of S1 to S10 and T1 to T10 of TCP

The energy gap between $T_2$ and $T_1$ of TCP is just 0.15 eV (Figure S17), which suggests that the high singlet formation ratio of TCP based OLED is not originated from the higher-level RISC mechanism.

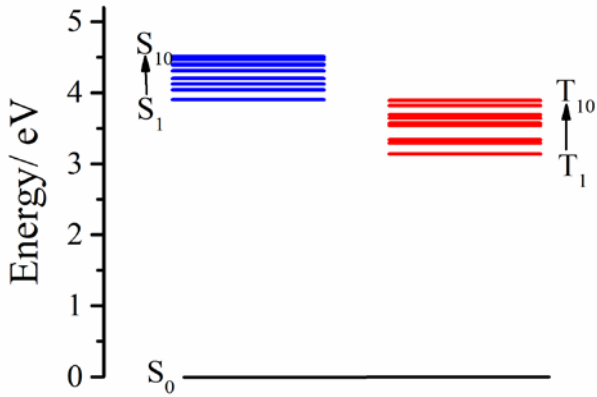

**Figure S17.** The calculated energy levels of $S_1$ to $S_{10}$ and $T_1$ to $T_{10}$ of TCP from time-dependent DFT calculation.

**SII-8. The MC measurement of TCP-based OLED**

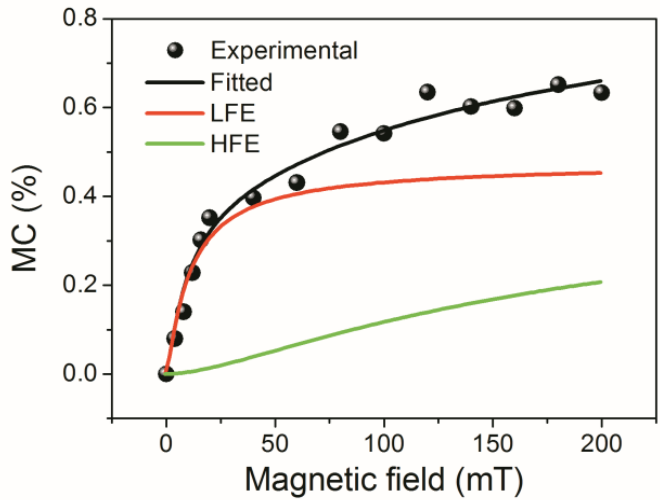

**Figure S18.** The MC as a function of the applied magnetic field (B) of the TCP based OLED at 8V, the MC can be fitted by the formula of $MC = \alpha B^2/(B+B_L)^2 + \beta B^2/(B+B_H)^2$, the first term is the low field effect (LFE) and the second term is the high field effect (HFE), the fitted parameters $\alpha$, $\beta$, $B_L$ and $B_H$ are 0.48, 0.47, 5 mT and 100 mT, respectively.

**SIII-1 The methods to calculate the TPI rate**

The triplet polaron interaction (TPI) rate can be obtained by the semi-classical Marcus theory, which reads,

$$k_{TPI} = \frac{1}{\hbar} V_{TPI}^2 \sqrt{\frac{\pi}{\lambda k_B T}} \exp\left[-\frac{(\Delta E_{ST} + \lambda)^2}{4\lambda k_B T}\right]$$

Where $V_{TPI}$ is the TPI electronic coupling; $\lambda$ is the total reorganization energy during the TPI process; $\Delta E_{ST}$ is the energy gap between $S_1$ and $T_1$ levels. For triplet-P$^-$ interaction, the hole hops from HOMO of triplet to HOMO of P$^-$, then $V_{TP^-I} \approx -\frac{\sqrt{3}}{2}\langle H_B|\hat{F}|H_A\rangle = -\frac{\sqrt{3}}{2}V_{hole}$ (see Scheme S3*a* and appendix); For triplet-P$^+$ interaction, the electron hops from LUMO of triplet to LUMO of P$^+$, then $V_{TP^+I} \approx -\frac{\sqrt{3}}{2}\langle L_B|\hat{F}|L_A\rangle = -\frac{\sqrt{3}}{2}V_{electron}$ (see Scheme S3*b* and appendix). $V_{hole}$ and $V_{electron}$ are the hole (electron) transfer integral. Z. An *et al.* have proposed other two-electron exchange mechanism for converting a triplet exciton into a singlet exciton via the TPI process (see Scheme S4 and appendix). [3] We also derive the electronic coupling formalism for this proposal (see appendix). Whereas the coupling includes only the two-electron integral term $-\frac{\sqrt{3}}{2}[(H_A L_B|H_A L_B) - (L_A L_B|L_A L_B)]$, which is quite small compared to the coupling mentioned above. Therefore, we think the proposed mechanism in this contribution is more significant.

(*a*) $T_1 + P^- \rightarrow P^- + S_1$  (*b*) $T_1 + P^+ \rightarrow P^+ + S_1$

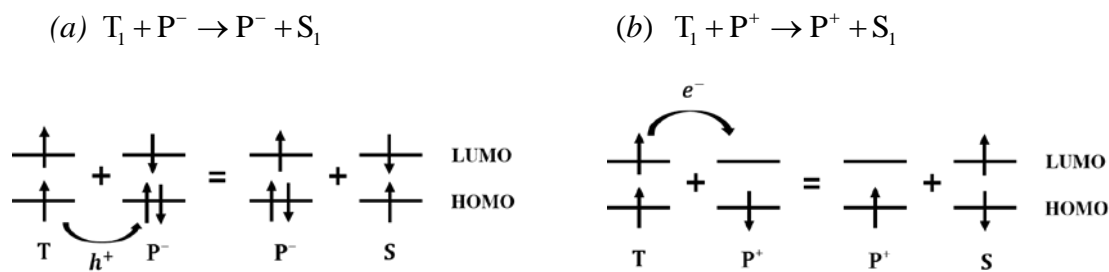

**Scheme S3** Schematic representation for hole (electron) hopping from HOMO (LUMO) of triplet to HOMO (LUMO) of P$^-$ (P$^+$).

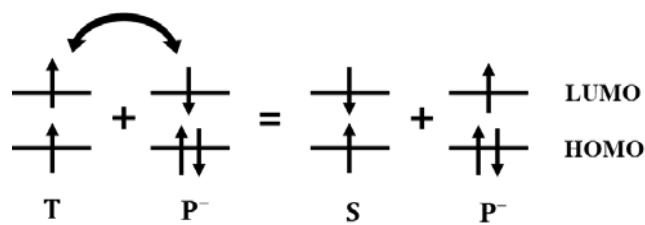

**Scheme S4** Two-electron exchange between LUMOs of triplet and P$^-$.

**Appendix**

When only consider the simple HOMO to LUMO excitation, the coupled dimer can be expressed as a doublet state with two $S_z$ components $|AB\rangle^{\frac{1}{2},\frac{1}{2}}$ and $|AB\rangle^{\frac{1}{2},-\frac{1}{2}}$. Using the Clebsch–Gordan coefficients to construct the spin adapted singlet, triplet and doublet states,

Considering the $|AB\rangle^{\frac{1}{2},\frac{1}{2}}$ state, the initial and final states are

$$|i\rangle = \sqrt{\frac{2}{3}}|A\rangle^{1,1}|B\rangle^{\frac{1}{2},-\frac{1}{2}} - \sqrt{\frac{1}{3}}|A\rangle^{1,0}|B\rangle^{\frac{1}{2},\frac{1}{2}} \tag{1}$$

$$|f\rangle = |A\rangle^{\frac{1}{2},\frac{1}{2}}|B\rangle^{0,0} \tag{2}$$

Considering the $|AB\rangle^{\frac{1}{2},-\frac{1}{2}}$ state, the initial and final states are

$$|i\rangle = -\sqrt{\frac{2}{3}}|A\rangle^{1,-1}|B\rangle^{\frac{1}{2},\frac{1}{2}} + \sqrt{\frac{1}{3}}|A\rangle^{1,0}|B\rangle^{\frac{1}{2},-\frac{1}{2}} \tag{3}$$

$$|f\rangle = |A\rangle^{\frac{1}{2},-\frac{1}{2}}|B\rangle^{0,0} \tag{4}$$

(I) For triplet-P$^-$ interaction process (Scheme 1a), *i.e.*, the hole transfer case, using the second quantization formalism, we get

$$|A\rangle^{1,1}|B\rangle^{\frac{1}{2},-\frac{1}{2}} = -a_{L\alpha}^{\dagger}a_{H\beta}b_{L\beta}^{\dagger}|HF_A HF_B\rangle \tag{5}$$

$$|A\rangle^{1,0}|B\rangle^{\frac{1}{2},\frac{1}{2}} = \frac{1}{\sqrt{2}}(a_{L\alpha}^{\dagger}a_{H\alpha} - a_{L\beta}^{\dagger}a_{H\beta})b_{L\alpha}^{\dagger}|HF_A HF_B\rangle \tag{6}$$

$$|A\rangle^{\frac{1}{2},\frac{1}{2}}|B\rangle^{0,0} = \frac{1}{\sqrt{2}}a_{L\alpha}^{\dagger}(b_{L\alpha}^{\dagger}b_{H\alpha} + b_{L\beta}^{\dagger}b_{H\beta})|HF_A HF_B\rangle \tag{7}$$

$$|A\rangle^{1,-1}|B\rangle^{\frac{1}{2},\frac{1}{2}} = a_{L\beta}^{\dagger}a_{H\alpha}b_{L\alpha}^{\dagger}|HF_A HF_B\rangle \tag{8}$$

$$|A\rangle^{1,0}|B\rangle^{\frac{1}{2},-\frac{1}{2}} = \frac{1}{\sqrt{2}}(a_{L\alpha}^{\dagger}a_{H\alpha} - a_{L\beta}^{\dagger}a_{H\beta})b_{L\beta}^{\dagger}|HF_A HF_B\rangle \tag{9}$$

$$|A\rangle^{\frac{1}{2},-\frac{1}{2}}|B\rangle^{0,0} = \frac{1}{\sqrt{2}}a_{L\beta}^{\dagger}(b_{L\alpha}^{\dagger}b_{H\alpha} + b_{L\beta}^{\dagger}b_{H\beta})|HF_A HF_B\rangle \tag{10}$$

where $\alpha, \beta$ denotes the electron spin, $L(H)$ indicates the HOMO (LUMO) orbital, and $a^{\dagger}(b^{\dagger}), a(b)$ is electron creation and annihilation in the two neighbour molecules A and B. $|HF_A HF_B\rangle$ represents the double occupied Hartree-Fock configuration.

Finally, the $S_z=1/2$ component shows the same electronic coupling between $|i\rangle$ and $|f\rangle$ as the $S_z = -1/2$ component, that is,

$$V_{TP^-I} = \langle i|\hat{H}|f\rangle = -\frac{\sqrt{3}}{2}\langle H_B|\hat{F}|H_A\rangle - \frac{\sqrt{3}}{2}[(H_A H_B|L_B L_B) + (H_A H_B|L_A L_A) - 2(L_B H_A|L_B H_B)]$$
$$\approx -\frac{\sqrt{3}}{2}\langle H_B|\hat{F}|H_A\rangle = -\frac{\sqrt{3}}{2}V_{hole} \quad (11)$$

where $(ab|cd)$ is chemists' notation of two-electron integral, and $\hat{F}$ is Fock operator.

(II) For triplet-$P^+$ interaction process (Scheme 1*b*), *i.e.*, the electron transfer case,

$$|A\rangle^{1,1}|B\rangle^{\frac{1}{2},-\frac{1}{2}} = -a^\dagger_{L\alpha}a_{H\beta}b_{H\alpha}|HF_A HF_B\rangle \quad (12)$$

$$|A\rangle^{1,0}|B\rangle^{\frac{1}{2},\frac{1}{2}} = -\frac{1}{\sqrt{2}}(a^\dagger_{L\alpha}a_{H\alpha} - a^\dagger_{L\beta}a_{H\beta})b_{H\beta}|HF_A HF_B\rangle \quad (13)$$

$$|A\rangle^{\frac{1}{2},\frac{1}{2}}|B\rangle^{0,0} = -\frac{1}{\sqrt{2}}a_{H\beta}(b^\dagger_{L\alpha}b_{H\alpha} + b^\dagger_{L\beta}b_{H\beta})|HF_A HF_B\rangle \quad (14)$$

$$|A\rangle^{1,-1}|B\rangle^{\frac{1}{2},\frac{1}{2}} = -a^\dagger_{L\beta}a_{H\alpha}b_{H\beta}|HF_A HF_B\rangle \quad (15)$$

$$|A\rangle^{1,0}|B\rangle^{\frac{1}{2},-\frac{1}{2}} = \frac{1}{\sqrt{2}}(a^\dagger_{L\alpha}a_{H\alpha} - a^\dagger_{L\beta}a_{H\beta})b_{H\alpha}|HF_A HF_B\rangle \quad (16)$$

$$|A\rangle^{\frac{1}{2},-\frac{1}{2}}|B\rangle^{0,0} = \frac{1}{\sqrt{2}}a_{H\alpha}(b^\dagger_{L\alpha}b_{H\alpha} + b^\dagger_{L\beta}b_{H\beta})|HF_A HF_B\rangle \quad (17)$$

Finally, the $S_z=1/2$ component shows the same electronic coupling between $|i\rangle$ and $|f\rangle$ as the $S_z = -1/2$ component, that is,

$$V_{TP^+I} = \langle i|\hat{H}|f\rangle = -\frac{\sqrt{3}}{2}\langle L_B|\hat{F}|L_A\rangle + \frac{\sqrt{3}}{2}[(L_A L_B|H_B H_B) + (L_A L_B|H_A H_A) - 2(L_A H_B|L_B H_B)]$$
$$\approx -\frac{\sqrt{3}}{2}\langle L_B|\hat{F}|L_A\rangle = -\frac{\sqrt{3}}{2}V_{electron} \quad (18)$$

(III) For the two-electron exchange case proposed by Z. An *et al.*, the coupling is

$$\langle i|\hat{H}|f\rangle = -\frac{\sqrt{3}}{2}[(H_A L_B|H_A L_B) - (L_A L_B|L_A L_B)] \quad (19)$$

**SIII-2 The computational details**

    The structures and energies for the crucial points in the adiabatic potential energy surface (see Scheme S5) were determined with DFT/TD-DFT theory implemented in the D.01 version of the Gaussian 09 package.[4] No symmetric constraint was adopted for all gas-phase calculations. Optimizations for the charged $P^+(P^-)$ and $T_1$ geometries were performed at the UB3LYP/6-31G[5,6] level. The $S_1$ geometry was optimized by using TD-RB3LYP/6-31G. The intermolecular transfer integral $\langle H_B|\hat{F}|H_A\rangle$ and $\langle L_B|\hat{F}|L_A\rangle$ were calculated with the

site-energy-corrected couplingmethod,[7] $H_{A(B)}$ and $L_{A(B)}$ are the frontier molecular orbital (HOMO and LUMO) of an isolated molecule A (B) in the dimer. PW91PW91/6-31G(d)[8] was employed to calculate the transfer integral according to the benchmark evaluations.[9]

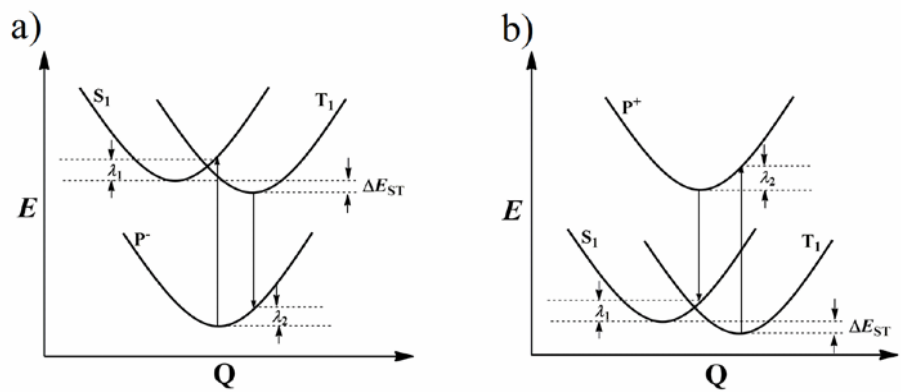

**Scheme S5** Schematic representation for calculation of the reorganization energy for TP$^-$I (*a*) and TP$^+$I (*b*) processes.

**SIII-3 The resluts of quantum chemistry computation**

**Table S4.** Calculated energy gap $\Delta E_{ST}$ (eV) and reorganization energies $\lambda_1$, $\lambda_2$, $\lambda$ (eV) at the (TD)-B3LYP/6-31Glevel.

|  | $\Delta E_{ST}$ | $\lambda_1$ | $\lambda_2$ | $\lambda$ |
|---|---|---|---|---|
| TP$^-$I | 0.437 | 0.066 | 0.313 | 0.379 |
| TP$^+$I | 0.437 | 0.305 | 0.290 | 0.595 |

**Table S5.** Transfer integrals for **TPA-TAZ** within 12 Å of the centroid molecule calculated at the PW91PW91/6-31G(d) level.

|  | Pathway | Distance (Å) | $V_{hole}$ (meV) | $V_{electron}$ (meV) |
|---|---|---|---|---|
| **TPA-TAZ** | 1 | 8.36 | **15.29** | 0.35 |
|  | 2 | 7.33 | 1.16 | **27.67** |
|  | 3 | 9.61 | 0.005 | 0.18 |
|  | 4 | 11.99 | 0.004 | 0.68 |
|  | 5 | 6.91 | 11.28 | 8.17 |
|  | 6 | 9.61 | 0.006 | 0.18 |

**Table S6.** Theoretical prediction of hole (electron) hopping rate $k_{TP^-I}$ ($k_{TP^+I}$) during the triplet-P$^-$ (triplet-P$^+$) interaction process at room temperature ($T = 300\ K$) for **TPA-TAZ** with Marcus theory.

|  | Pathway | $k_{TP^-I}$ (s$^{-1}$) | $k_{TP^+I}$ (s$^{-1}$) |
|---|---|---|---|
| **TPA-TAZ** | 1 | **2.00×10$^5$** | 5.99×10$^1$ |
|  | 2 | 1.15×10$^3$ | **3.79×10$^5$** |
|  | 3 | 2.13×10$^{-2}$ | 1.60×10$^1$ |
|  | 4 | 1.37×10$^{-2}$ | 2.28×10$^2$ |
|  | 5 | 1.09×10$^5$ | 3.30×10$^4$ |
|  | 6 | 3.07×10$^{-2}$ | 1.62×10$^1$ |

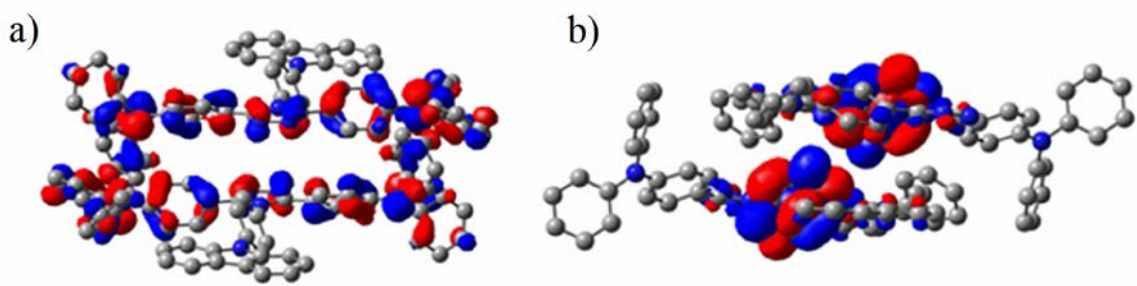

**Figure S19.** Molecular packing motifs of HOMOs (*a*) and LUMOs (*b*) for the most effective pathway of **TPA-TAZ**. The hydrogen atoms are omitted for clarity.